\documentclass{aa}  

\usepackage{graphicx}
\usepackage{txfonts}
\usepackage{natbib}
\usepackage{subfigure}
\usepackage{ulem}

\usepackage{hyperref}
\hypersetup{
 colorlinks=true,
 citecolor=blue,
 linkcolor=blue,
 pdftitle={ Eccnetric MBHBs in PTA -- Fisher Matrix },
 }

\usepackage{color}
\usepackage[dvipsnames]{xcolor}
\definecolor{myorange}{rgb}{0.8, 0.3, 0.0}

\begin{document}

   %\title{Resolving the nano-Hz gravitational wave sky}
   \title{Resolving the nano-Hz gravitational wave sky: the detectability of eccentric binaries with PTA experiments}
    \titlerunning{Detectability of eccentric binaries with PTA experiments}
   \subtitle{}

   \author{Riccardo J. Truant\inst{1}\fnmsep\thanks{r.truant@campus.unimib.it} \and David Izquierdo-Villalba\inst{1,2} \and Alberto Sesana\inst{1,2} \and \\  Golam Mohiuddin Shaifullah\inst{1,2} \and Matteo Bonetti\inst{1,2,3}
          % \and
          % C. Ptolemy\inst{2}\fnmsep\thanks{Just to show the usage
          % of the elements in the author field}
          }

            \institute{Dipartimento di Fisica ``G. Occhialini'', Universit\`{a} degli Studi di Milano-Bicocca, Piazza della Scienza 3, I-20126 Milano, Italy
              \and
               INFN, Sezione di Milano-Bicocca, Piazza della Scienza 3, 20126 Milano, Italy
               \and
               INAF - Osservatorio Astronomico di Brera, via Brera 20, I-20121 Milano, Italy
              % \email{}
         % \and
             % University of Alexandria, Department of Geography, ...\\
             % \email{c.ptolemy@hipparch.uheaven.space}
             % \thanks{The university of heaven temporarily does not
             %         accept e-mails}
             }

   \date{Received ---; accepted ---}
   
   \abstract{
   
   % Arxiv only allows 1920 characters in the abstract.

   % Character number 1917:
   Pulsar Timing Array (PTA) collaborations reported evidence of a nano-Hz stochastic gravitational wave background (sGWB) compatible with an adiabatically inspiraling population of massive black hole binaries (MBHBs). Despite the large uncertainties, the relatively flat spectral slope of the recovered signal suggests a possible prominent role of MBHB dynamical coupling with the environment or/and the presence of an eccentric MBHB population. This work aims at studying the capabilities of future PTA experiments to detect single MBHBs under the realistic assumption that the sGWB is originated from an eccentric binary population coupled with its environment. To this end, we generalize the standard signal-to-noise ratio (SNR) and Fisher Information Matrix calculations used in PTA for circular MBHBs to the case of eccentric systems. We consider an ideal 10-year MeerKAT and 30-year SKA PTAs and apply our method over a wide number of simulated eccentric MBHB populations. We find that the number of resolvable MBHBs for the SKA (MeerKAT) PTA is ${\sim}\,30$ ($4$) at $\rm SNR\,{>}\,5$ (${>}\,3$), featuring an increasing trend for larger eccentricity values of the MBHB population. This is the result of eccentric MBHBs at ${\lesssim}\,10^{-9}\, \rm Hz$ emitting part of their power at high harmonics, thus reaching the PTA sensitivity band. Our results also indicate that resolved MBHBs do not follow the eccentricity distribution of the underlying MBHB population, but prefer low eccentricity values (${<}\,0.6$). Finally, the recovery of binary intrinsic properties and sky-localization do not depend on the system eccentricity, while orbital parameters such as eccentricity and initial orbital phase show clear trends. Although simplified, our results show that SKA will enable the detection of tens of MBHBs, projecting us into the era of precision gravitational wave astronomy at nano-Hz frequencies.

   }

   \keywords{ black hole physics – gravitational waves  – methods: numerical – methods: statistical }

   \maketitle

%-------------------------------------------------------------------------------------------------------------------------------------------
%-------------------------------------------------------------------------------------------------------------------------------------------
\section{Introduction}
In the last three decades, multi-wavelength observations have pointed out that massive black holes ($\rm {>}\,10^6\,M_{\odot}$, MBHs) reside at the centre of most of the galaxies, co-evolving with them and powering quasars and active galactic nuclei \citep{Schmidt1963,Genzel1987,Kormendy1988a,Dressler1988,Kormendy1992,Haehnelt1993,Genzel1994,Faber1999,ODowd2002,HaringRix2004,Peterson2004,Vestergaard2006,Hopkins2007,MerloniANDHeinz2008,Kormendy2013,Ueda2014,Savorgnan2016}. Galaxies do not evolve in isolation and, in the context of the currently favored hierarchical clustering scenario for structure formation, they are expected to merge frequently \citep{WhiteandRees1978,WhiteFrenk1991,Lacey1993}. Consequently, the presence of MBHs lurking in the centers of galaxies and the important role of galactic mergers suggest that massive black hole binaries (MBHBs) have formed and coalesced throughout cosmic history.\\

The dynamical evolution of MBHBs is ruled by many different processes \citep[see e.g. the seminal work by][]{Begelman1980}. Following the merger of the two parent galaxies, dynamical friction, exerted by dark matter stars and gas, drags the two MBHs towards the nucleus of the newly formed system, reducing the initial MBH separation ($\rm {\sim}\,kpc$ scales) dawn to a few parsecs \citep{Yu2002,Mayer2007,Fiacconi2013,Bortolas2020,Bortolas2022,Kunyang2022}. At these distances, a bound binary forms and dynamical friction ceases to be efficient. Interactions with single stars or torques extracted from a circumbinary gaseous disc take the main role in further evolving the MBHB separation \citep{Quinlan1997,Sesana2006,Vasiliev2014,Sesana2015,Escala2004,Dotti2007,Cuadra2009,Bonetti2020,Franchini2021,Franchini2022}. These processes harden the MBHB down to sub-pc scales, where the emission of gravitational waves (GWs) drives it to final coalescence. During this last evolutionary stage, MBHBs are powerful GW sources, whose emission spans over a wide range of frequencies. In particular,  low-$z$, high mass ($\rm {>}\,10^7\,M_{\odot}$) inpiralling MBHBs emit GWs in the nano-Hz frequency window ($10^{-9}\,{-}\,10^{-7}$ Hz), probed by Pulsar Timing Array (PTA) experiments \citep{Foster1990}. 

By monitoring an array of millisecond pulsars and measuring the changes in the time-of-arrival of their pulses, PTAs are sensitive to the incoherent superposition of all the GWs coming from the cosmic population of MBHBs \citep{Sazhin1978,Detweiler1979}. The overall signal is thus expected to have the properties of a stochastic GW background (sGWB). The specific amplitude and spectral shape of the signal are closely related to the galaxy merger rate and the environment in which MBHBs shrink \citep{Phinney2001,Jaffe2003,2011MNRAS.411.1467K,2013CQGra..30v4014S,2014MNRAS.442...56R}, and it can be disentangled from other stochastic noise processes affecting PTA measurements thanks to its distinctive correlation properties \citep{HellingsAndDowns1983}. Moreover, because of the sparseness of the most massive and nearby binaries, individual deterministic signals, usually referred to as continuous GWs (CGWs) might also be resolved \citep{Sesana2009,Rosado2015,Agazie2023_CGW,Antoniadis2023_CGW}. Those would provide precious information about the most massive and nearby MBHBs in the universe and are ideal targets to extend multimessenger astronomy in the nano-Hz GW band \citep[e.g.][]{2013CQGra..30v4013B}. For this reason, both types of signals (CGW and sGWB) are of great interest for PTA observations.

There is currently several operational PTA collaborations around the world: the \textit{European Pulsar Timing Array} \citep[EPTA,][]{Kramer2013,Desvignes2016}, the \textit{North American Nanohertz Observatory for Gravitational Waves} \citep[NANOGrav,][]{McLaughlin2013,Arzoumanian2015}, the \textit{Parkes Pulsar Timing Array} \citep[PPTA,][]{Manchester2013,Reardon2016}, the \textit{Indian PTA} \citep[InPTA,][]{Susobhanan2021}, the \textit{Chinese PTA} \citep[CPTA,][]{Lee2016} and the MeerKAT PTA \citep[MPTA,][]{Miles2023}. The latest results published by several of those collaborations report evidence about the presence of an sGWB \citep[at $2\,{-}\,4\sigma$ significance level,][]{Antoniadis2023,Agazie2023,Reardon2023,Xu2023}, compatible with the existence of low-$z$ MBHBs \citep{Afzal2023,InterpretationPaperEPTA2023}. 
Quite interestingly, the amplitude of the signal is at the upper end of the predicted range of MBHB populations \citep[see e.g.][]{Sesana2013}, and the best fit to the logarithmic spectral slope of the signal appears to deviate from the vanilla $-2/3$ value, expected from a circular population of MBHBs evolving solely through GW emission. Although uncertainties are too large to draw any conclusion, these two facts might bear important implications for the underlying MBHB population. On the one hand, the high signal amplitude likely implies a large contribution from very massive binaries at the upper-end of the MBH mass function. On the other hand, the tentative deviation in the spectral slope can hint at a strong coupling of the binaries with their stellar environment \citep[i.e. stellar hardening,][]{Quinlan1997,Sesana2006} or at non-negligible orbital eccentricities \citep[see e.g][]{Gualandris2022,Fastidio2024}.

In light of the above considerations, it is therefore interesting to investigate the expected statistical properties of CGWs that might be resolved by future PTA experiments. In fact, although the topic has been addressed by several authors \citep[e.g.][]{Rosado2015,Kelley2018,Gardiner2023}, most of the current literature focuses on fairly idealised cases. Both \cite{Rosado2015} and \cite{Gardiner2023} assume circular binaries, and a vast range of models that are not necessarily tailored to the currently detected signal. \cite{Kelley2018} brings eccentricity in the picture, but employs a very simplified description of the signal and investigates a scenario where the GWB is almost a factor of three smaller than what currently inferred from the data. Moreover, none of the work above touches on PTA capabilities to estimate the source parameters. Work on this subject has so far involved only circular binaries \cite[e.g.][]{SesanaVecchio2010,Ellis2012,Goldstein2019} and when eccentricity has been considered \citep{2016ApJ...817...70T}, results have never been scaled at the overall MBHB population level.

%....(TO BE CONTINUED) \mb{???}
%Assuming that in the upcoming years the latest results of PTA collaborations will acquire significance, future searches for CGWs should take into account the environmental coupling and the MBHB eccentricity, with the standard methodology for CGW detection eventually moving away from the canonical circular case and framed in the generic scenario of eccentric MBHBs. To date, several authors have explored the possibility of detecting CGW sources employing idealized PTA experiments, but they mostly focused on circular MBHBs \citep{Rosado2015,Kelley2018,Gardiner2023}, or made simplistic assumptions in the treatment of eccentric CGWs \citep{Kelley2018}.
%are circular and purely evolving through GW emission, neglecting (i) the effect of eccentricity in the sGWB and (ii) the fact that the CGW source spreads its signal along different harmonic. 

In this work, we aim to relax several of the assumptions made in previous investigations with the goal of providing an extensive assessment of future PTA experiments capabilities of resolving CGWs. To this end, we employ state-of-the-art MBHB populations including environmental coupling and eccentricity, tailored to reproduce the observed PTA signal. We also adapt the formalism of \cite{2004PhRvD..69h2005B} to PTA sources to develop a fast Fisher Information Matrix algorithm for eccentric binaries, limiting the description of the signal to the Earth term only. We apply this machinery to putative 10-year MeerKAT and 30-year SKA PTAs, and we identify resolvable CGWs via iterative subtraction \citep{2021PhRvD.104d3019K} over a wide range of simulated eccentric MBHB populations compatible with the latest amplitude of the sGWB, offering a realistic assessment of the future potential of PTAs.

%relax these assumptions and explore the capabilities of detecting the CGW from single sources when these are eccentric and embedded in an sGWB generated by an eccentric MBHB population. To this end, we generalized the standard framework for circular MBHB search used in PTA and we derived the computation of the signal-to-noise ratios (SNR) and Fisher Information Matrix for eccentric cases. We have adopted a 10-year MeerKAT and 30-year SKA PTAs and applied our search over a wide number of simulated eccentric MBHB populations compatible with the latest amplitude of the sGWB.\\

The paper is organized as follows. In Section~\ref{sec:Methodology} we overview the methodology used to characterize the emission of eccentric MBHBs and the time residuals that they imprint in PTA data. In Section~\ref{sec:DetectionSMBHBs}, we describe the computation of the signal-to-noise ratio and Fisher Information Matrix for eccentric MBHBs. In Section~\ref{sec:pop and PTA}, we present the population of eccentric MBHBs and the PTA experiments that we use. In Section~\ref{sec:Results}, we discuss the results, focusing on the number of resolvable sources and the effect of the eccentricity in determining their number and their parameter estimation. In Section~\ref{sec:Caveats}, we discuss some of the caveats of the present implementation and finally, in Section~\ref{sec:Conclusions}, we summarise the main results of the paper. A Lambda Cold Dark Matter $(\Lambda$CDM) cosmology with parameters $\Omega_{\rm m} \,{=}\,0.315$, $\Omega_{\rm \Lambda}\,{=}\,0.685$, $\Omega_{\rm b}\,{=}\,0.045$, $\sigma_{8}\,{=}\,0.9$ and $h \, {=} \, \rm H_0/100\,{=}\,67.3/100\, \rm km\,s^{-1}\,Mpc^{-1}$ is adopted throughout the paper \citep{PlanckCollaboration2014}.

%{\color{red} This is not totally true, I'll re-wrte it: It could be interpreted as a hint that the evolutionary pace of MBHB at nano-Hz is not only driven by GW emission but also by the interaction with the galactic environment (gas or stars). According to galaxy evolution theory, PTA MBHB sources are expected to be embedded in gas-poor environments, being stellar interactions the main environmental mechanisms ruling the orbital evolution. These kinds of interactions cause the eccentricity of the MBHB system to rise \citep{Quinlan1997,Sesana2006}. Consequently, an environmental coupling can yield a fraction of the nano-Hz MBHBs to exhibit eccentric orbits which spread their emitted GW signal towards different harmonic. The net result is an attenuation of the sGWB signal at low frequencies and a mild boost at higher ones \citep[see e.g][]{Bonetti2023}.}\\
\section{The gravitational emission of eccentric supermassive black holes binaries} \label{sec:Methodology}
In this section, we outline the basic concepts used to explore the detectability of CGWs generated by eccentric MBHBs.%, typically $\sim 10^{-9}-10^{-7}$ Hz. %A GW passing across the line of sight between the Earth perturbs space-time affecting the time periodicity of the pulse as seen by the Earth. In the following section, we present GW signal in PTA experiments \textbf{Improve this part}

%\subsection{Add short introduction on Kepler problem and orbit decomposition}
%\davcoment{Not sure if it will be needed...}

\subsection{The gravitational wave signal} \label{sec:GWSignal}

The GW metric perturbation $h_{ab}(t)$ in the trace-less and transverse (TT) gauge can be written as a linear superposition of two polarizations ($h_{+}$ and $h_{\times}$) and their base tensor ($e^{+}_{ab}$ and $e^{\times}_{ab}$):
\begin{equation}
    h_{ab}(t,\hat{\Omega}) = h_+(t)e^+_{ab} (\hat{\Omega})+h_\times(t) e^\times_{ab} (\hat{\Omega})
\end{equation}
where $\hat{\Omega}$ is the GW propagation direction. Differently from the monochromatic emission of a circular binary, the GW signal of an eccentric source is spread over a spectrum of harmonics of the orbital frequency. To model the signal of these eccentric sources, we use the GW waveform presented in \cite{2016ApJ...817...70T}, which provides the analytic expression of the GW emission of an eccentric binary. Specifically, adopting the quadrupolar approximation and making use of the Fourier analysis of the Kepler problem, $h_{+,\times}$ can be written  as:
\begin{equation} \label{eq:h_plus_cross}
    \begin{aligned}
        h_+(t) &\,{=}\,  \sum_{n = 1}^{\infty} -\left(1+\cos^2{\iota}\right) \left[a_n\cos{(2\gamma)} - b_n\sin{(2\gamma)}\right] + \left(1-\cos^2{\iota}\right)c_n,\\ 
        %h_+(t) &=  \sum_{n = 1}^{n = \infty} -\left(1+\cos^2{\iota}\right) \left[a_n\cos{(2\gamma)} - b_n\sin{(2\gamma)}\right] + \left(1-\cos^2{\iota}\right)c_n,\\ 
        h_\times(t) &\,{=}\, \sum_{n = 1}^{\infty} 2\cos{\iota}\left[b_n \cos{(2\gamma)} + a_n \sin{(2\gamma)}\right],
    \end{aligned}
\end{equation}
where
\begin{equation}
    \begin{aligned}
        a_n \,{=}\,  &- n\zeta\omega^{2/3}J_{n-2}(ne) -2eJ_{n-1}(ne)+(2/n) J_n(ne) \\%[-1ex] 
              &+ 2eJ_{n+1}(ne)-J_{n+2}(ne)\cos{(nl(t))}, \\
        b_n \,{=}\, &-n\zeta\omega^{2/3} \sqrt{1-e^2}\left[J_{n-2}(ne)-2J_n(ne) + J_{n+2}(ne)\right]\sin{(nl(t))}, \\
        c_n \,{=}\, &2\zeta\omega^{2/3}J_n(ne)\cos(nl(t)).
    \end{aligned}
\end{equation}
being $e$ the eccentricity of the MBHB, $n$ the harmonic number and $J_n(x)$ the $n{-}\rm th$ Bessel Function of the first kind. $\zeta$ is the GW amplitude given by the combination of the redshifted chirp mass, $\mathcal{M}_z$, and the luminosity distance, $D_L$:
\begin{equation}
\zeta \,{=}\, \dfrac{(G\mathcal{M}_z)^{5/3}}{c^4D_L}.  
\end{equation}
Here, $c$ is the light speed and $G$ is the gravitational constant. The redshifted chirp mass is expressed as:
\begin{equation}
    \mathcal{M}_{z}\,{=}\,\mathcal{M} (1+z)\,{=}\, \dfrac{M q^{5/3}}{(1+q)^{6/5}}(1+z),
\end{equation}
being $\mathcal{M}$ the rest-frame chirp mass, $M\,{=}\, m_{1} \,{+}\, m_{2}$ the total mass of the binary in the rest frame, and $q\,{=}\,m_{2}/m_{1}\,{<}\,1$ its mass ratio. With this definition of $q$, $m_{1}$ and $m_{2}$ are identified as the mass of the primary and secondary MBH, respectively. The variable $\iota$ is the inclination angle defined as the angle between the GW propagation direction and the binary orbital angular momentum, $\hat{L}$. %, so that $\cos(\iota)\,{=}\,-\hat{L} \cdot \hat{\Omega} $. 
The quantity $l(t)$ refers to the binary mean anomaly $l(t)\,{=}\,l_0 + 2\pi \int_{t_0}^t f_k(t') dt' $, where$f_k(t')$ corresponds to the observed Keplerian frequency defined as $f_k(t)\,{=}\,(1+z) f_{k,r}(t)$ with $f_{\mathrm{k},r}(t)\,{=}\, (2\pi)^{-1}  \sqrt{GM/r_{bin}(t)^3}$ the rest frame Keplerian frequency and $r_{bin}(t)$ the semi-major axis of the MBHB orbit. In principle, $r_{bin}(t)$ can evolve during the observation time due to the GW emission and environmental interaction. However, here we assume that the orbital frequency and eccentricity do not evolve during the observation time and, hence, $f_k(t)$ can be treated as a constant value, i.e $f_k$ (see discussion about this assumption in Section~\ref{sec:Caveats}). The orbital angular frequency is then given by $\omega\,{=}\,2\pi f_k$, while  $\gamma$ is the angle that measures the direction of the pericenter with respect to the direction $\hat{x}$, defined as $\hat{x} \equiv (\hat{\Omega} + \hat{L}\cos{i})/\sqrt{1-\cos^2{i}}$.\\ %Finally, $e $ is the orbit eccentricity and $J_{(.)}(.)$ are the Bessel functions. 

\subsection{The timing residuals} \label{sec:TimeResiduals}
A GW passing between the pulsar and the Earth perturbs the space-time metric, causing a modification in the arrival time of the pulse to the Earth. This induces a fractional shift in the pulsar rotational frequency, $z(t,\hat{\Omega})$, given by \citep[e.g.][]{Anholm2009,Book2011}:
\begin{equation}
\begin{aligned} \label{eq:GW redshift}
    z(t,\hat{\Omega}) = \dfrac{1}{2}\dfrac{p^ap^b}{1+\hat{\Omega} \cdot \hat{p}} \Delta h_{ab},
\end{aligned}
\end{equation}
where $\hat{p}$ is the unit direction vector to the pulsar and $\Delta h_{ab}=h_{ab}(t,\vec{x}_{E}) - h_{ab}(t_P,\vec{x}_{P})$ is the difference in the metric perturbation computed at the moment in which the GW arrives at the solar system barycenter ($t$), and when it passed through the pulsar ($t_P=t-L/c$, where $L$ denotes the distance to the pulsar). We define $\vec{x}_{E}$ to coincide with the solar system barycenter, which is the origin of the adopted coordinate system, while $\vec{x}_{P}\,{=}\, L \hat{p}$ corresponds to the pulsar sky position. %PTA experiments measure the offset between the expected pulse time of arrival (ToA), constructed according to a deterministic model, and the raw ToA which is influenced by the passage of GWs and all the unmodelled phenomena. 
In practice, what PTA experiments measure is the \textit{timing residual}, which corresponds to the time integrated effect of Eq.~\eqref{eq:GW redshift}: %from the start of the observing campaign to the epoch of the considered pulsar observation ($T_{\rm obs}$):
\begin{equation} \label{Eq::time res}
    \begin{aligned}
    s(t) = &\int_{t_0}^{t} dt' z(t') \\
           & =F^+(\hat{\Omega})\left[s_+(t)-s_+(t_0)\right] + F^\times(\hat{\Omega})\left[s_\times(t)-s_\times(t_0)\right].
    \end{aligned}
\end{equation}
Here, $s_{+,\times}(t) \,{=}\,\int_{t_0}^t h_{+,\times}(t') dt'$, $t_0$ is the time at which the observational campaign starts and $t\,{=}\,t_0 + T_{\rm obs}$ is the epoch of the considered observation. The variable $F^{+,\times}$ denote the \textit{antenna pattern functions} and encode the geometrical properties of the detector (for a PTA experiment the test masses are the Earth and the pulsar). In particular, $F^{+,\times}$ depends on the GW propagation direction ($\hat{\Omega}$) and the pulsar sky location ($\hat{p}$). By making use of the polarization basis tensor $\{\hat{n},\hat{u},\hat{v}\}$ (see \citealt{2021arXiv210513270T}), the pattern functions $F^{+,\times}$ can be can be written as:
\begin{equation}
    F^+(\hat{\Omega}) \,{=}\, \dfrac{1}{2} \dfrac{[\hat{u}\cdot\hat{p}]^2 -[\hat{v}\cdot\hat{p}]^2}{1+\hat{\Omega}\cdot\hat{p}},
\end{equation}
and
\begin{equation}
    F^\times(\hat{\Omega}) \,{=}\, \dfrac{\bigl[\hat{u}\cdot\hat{p}\bigr][\hat{v}\cdot\hat{p}]}{1+\hat{\Omega}\cdot\hat{p}},
\end{equation}
where $\hat{n}$ corresponds to the vector pointing to the GW source:
\begin{equation}
    \hat{n} \,{=}\, -\hat{\Omega} = [\cos{\theta}\cos{\phi} ,\, \cos{\theta}\sin{\phi}, \, \sin{\theta} ],
\end{equation}
and $\hat{u}$ and $\hat{v}$ are defined as: %$\hat{u}=(\hat{n}\times\hat{L})/\lvert \hat{n}\times\hat{L} \rvert$ and $\hat{v}= \hat{u}\times n$ such that:
\begin{equation}
    \begin{aligned}
     \hat{u} \,{=}\, \frac{\hat{n}\times\hat{L}}{\lvert \hat{n}\times\hat{L} \rvert} & \,{=}\, [\cos{\psi}\sin{\theta}\cos{\phi} -\sin{\psi}\cos{\theta},  \\%[-1ex]
             & \quad \cos{\psi}\sin{\theta}\sin{\phi} + \sin{\psi}\cos{\phi}, \\%[-1ex]
             & \quad -\cos\psi\cos\theta],
    \end{aligned}
\end{equation}

\begin{equation}
    \begin{aligned}
     \hat{v} \,{=}\, \hat{u}\times \hat{n} & = [ \sin{\psi}\sin{\theta}\cos{\phi} + \cos{\psi}\sin{\phi} ,    \\%[-1ex]
             &\quad  \sin{\psi}\sin{\theta}\sin{\phi} -\cos{\psi}\cos{\phi} , \\%[-1ex]
             &\quad  -\sin{\psi}\cos{\theta} ].
    \end{aligned}
\end{equation}
Here $\theta$ and $\phi$ are the sky location of the MBHB expressed in spherical polar coordinates $(\theta,\phi) \,{=}\, (\pi/2 -\text{DEC},\text{RA})$, being DEC the declination and RA the right ascension of the binary. Finally, $\psi$ is the polarization angle, ranging between $[0,\pi]$.\\

For simplicity, we consider only the Earth term\footnote{For further information about the impact of including the pulsar distance in the computation of the SNR, we refer to \citealp{2016ApJ...817...70T}.}, we ignore any time evolution of the MBHB frequency and neglect higher-order post-Newtonian effects such as the pericenter precession and orbit-spin coupling. Those are expected to play a minor role in the output of PTA GW signal \citep{SesanaVecchio2010}, and can be safely neglected, at least for a first order estimate. We will comment on the validity of these assumptions in Section~\ref{sec:Caveats}. Under  the framework outlined above, the values of the timing residuals can be written analytically as \citep{2016ApJ...817...70T}:
\begin{equation} \label{eq:splus_scross}
    \begin{aligned}
        s_+(t) &= \sum_{n=1}^{\infty} -\left(1+\cos^2{i}\right)\left[\tilde{a}_n\cos{(2\gamma)} - \tilde{b}_n\sin(2\gamma)\right] \\
        &+ \left(1-\cos^2{i}\right)\tilde{c}_n, \\
        s_\times(t) & = \sum_{n=1}^{\infty} 2\cos{i} \,\left[\tilde{b}_n\cos{(2\gamma)}+\tilde{a}_n\sin{(2\gamma)} \right],
    \end{aligned}
\end{equation}
being:
\begin{equation}
    \begin{aligned}
        \tilde{a}_n = &-\zeta \omega^{-1/3}[J_{n-2}(ne)-2eJ_{n-1}(ne) (2/n)J_{n}(ne) \\%[-1ex] 
                    &+ 2eJ_{n+1}(ne) -J_{n+2}(ne) ]\sin(nl(t)) \\
        \tilde{b}_n = & \zeta \omega^{-1/3} \sqrt{1-e^2}[J_{n-2}(ne)-2J_{n}(ne)+J_{n+2}(ne)]\cos(nl(t))\\
        \tilde{c}_n = & (2/n) \zeta\omega^{-1/3}J_{n}(ne)\sin(nl(t)).
    \end{aligned}
\end{equation}

\section{SNR and parameter estimation of supermassive black hole binaries in PTA data} \label{sec:DetectionSMBHBs}
In this section, we introduce the methodology used to compute the \textit{signal-to-noise ratio} (hereafter, SNR) and the \textit{Fisher information matrix} from the gravitational wave signal emitted by a single eccentric MBHB.

\subsection{Signal-to-noise ratio for eccentric binaries}  \label{sec:SNR}

%The GW signal, in PTA dataset, emitted by an MBHB is generally buried inside the detector noise. 
In general, to assess the possibility of detecting a nano-Hz CGW signal generated by an MBHB it is required to determine how its signal compares with the background noise present in the detector. This is usually done by computing the SNR. Given the deterministic nature of the CGW signal, the optimal way to compute the SNR is through \textit{matched filtering}. Assuming that a CGW is present in the timing residual of a pulsar, the match filtering procedure gives the expression:
\begin{equation} \label{eq:SNR_Geneal}
    \left(\dfrac{S}{N}\right)^2=4 \int_{0}^{\infty} df \dfrac{\lvert {\tilde{s}}(f) \rvert ^2}{S_k(f)}.
\end{equation} 
Estimating the SNR therefore requires the characterization of the noise properties encoded in $S_k(f)$, i.e the noise power spectral density (NPSD) of the $k{-}th$ pulsar inside our array, besides the knowledge of the signal $\tilde{s}(f)$, which is the Fourier transforms of $s(t)$ given by Eq.~\eqref{eq:splus_scross}. \\ %\as{I tried to refer the equation for $s$ written above here, but somehow it doesn't work. The equation in fact is not numbered, I checked it but I cannot spot what's wrong, can you please check?} \davcoment{Done! ;)}.\\% Following the standard approach, here we assume that the noise is a Gaussian and zero-mean stochastic stationary process.\\ 

As shown by Eq.~\eqref{eq:SNR_Geneal}, the computation of the SNR requires the time residuals in the frequency domain, i.e. $\tilde{s}(f)$. However, as described in Section~\ref{sec:GWSignal}, in PTA experiments the time residuals are framed on the time domain. Transforming these in the frequency domain can be easily addressed in the case of circular binaries given that the CGW signal is monochromatic % and does not evolve during the observation time 
\citep[see e.g,][]{Rosado2015}. However, in the generic case of an eccentric binary, the signal is spread over a spectrum of harmonics and the term $\lvert {\tilde{s}}(f) \rvert ^2$ contains mixed products between residuals originated at different harmonics. %\riccoment{Under these circumstances, no analytical expression can be derived, NOT SO SURE}. 
To address this scenario we worked under the assumption that the noise is a Gaussian and zero-mean stochastic stationary process and adopted a similar approach to the one presented in \cite{2004PhRvD..69h2005B}. In brief, even if Eq.~\eqref{eq:SNR_Geneal} is given by the mixed product generated by different harmonics, their signal in the frequency domain is described by a delta function centred at the emission frequency  $nf_k$. Consequently, %once integrated over all the possible frequencies, 
the product of the residuals generated by harmonics emitting at different frequencies are orthogonal and cancel out. We can therefore treat each harmonic separately as a monochromatic signal and compute its SNR by exploiting the fact that
%Hence, when computing the contribution of each harmonic to the SNR in Eq.~\eqref{eq:SNR_Geneal}, the noise can be treated as it was white, allowing us to bring it out from the integral. Using the the Parseval theorem we then obtain: 
%To overcome this problem we worked under the assumption that the noise is a Gaussian and zero-mean stochastic stationary process and adopted a similar approach to the one presented in \cite{2004PhRvD..69h2005B}. In brief, even if Eq.~\eqref{eq:SNR_Geneal} is given by the mixed product generated by different harmonics, their signal in the frequency domain is described by a delta function centred at the emission frequency  $nf_k$. Consequently, %once integrated over all the possible frequencies, the product of the residuals generated by harmonics emitting at different frequencies are orthogonal and cancel out. Hence, when computing the contribution of each harmonic to the SNR in Eq.~\eqref{eq:SNR_Geneal}, the noise can be treated as it was white, allowing us to bring it out from the integral. Using the the Parseval theorem we then obtain: 
\begin{equation} \label{eq:innerproduct_f_to_t}
    \dfrac{4}{S_k(f_n)} \int_0^\infty \lvert \tilde{s_n}(f)\rvert^2 df \simeq \dfrac{2}{S_k(f_n)} \int_0^\infty s_n(t)^2 dt.
\end{equation}
The SNR from an eccentric MBHB, for a single pulsar, is thus given by the summation in quadrature over all the harmonics:
\begin{equation} \label{eq:SNR_ecc}
    \left(\dfrac{S}{N}\right)^2\,{=}\, \sum_{n=1}^{\infty} \dfrac{2}{S_k(f_n)} \int_{t_0}^t  dt'\, s_n^2(t'),
\end{equation}
while the total SNR in the PTA is given by the sum in quadrature of the SNRs produced in all the $N_{\rm Pulars}$ included in the array:
\begin{equation} \label{eq:SNR_tot_ecc}
    \left(\dfrac{S}{N}\right)^2_{tot} \,{=}\, \sum_{k=1}^{N_{\rm Pulars}} \left(\dfrac{S}{N}\right)^2_k.
\end{equation}
%where $N_{\rm Pulars}$ corresponds to the total number of pulsars in the PTA experiment.\\

Finally, the computation of the SNR of Eq.~\eqref{eq:SNR_ecc} requires a summation over all the harmonics, i.e $n \in [1,+\infty]$. However the contribution to the SNR goes to zero for $n\rightarrow +\infty$ and the sum can be appropriately truncated. To select the harmonic of truncation, $n_{\rm max}$, we adopted the simple criteria of $n_{\rm max} = 4 \, n_{\rm peak}$, being $n_{\rm peak}$ the harmonic number at which the power of the GW emission is maximized for the selected eccentricity. To compute this value we follow the numerical fit presented in \cite{2021RNAAS...5..275H}:
\begin{equation} \label{eq:nmax}
    n_{\text {peak }}(e) \simeq 2\left(1+\sum_{k=1}^4 c_k e^k\right)\left(1-e^2\right)^{-3 / 2} ,
\end{equation}
where $c_1\,{=}\,-1.01678$, $c_2\,{=}\,5.57372$, $c_3\,{=}\,-4.9271$, $c_4 \,{=}\, 1.68506$. We have checked how the exact value of $n_{\rm max}$ affects our results. Specifically, less than $1\%$ relative difference is seen in the SNR when it is computed assuming  $n_{\rm max} \,{=}\, 10^4 $ instead of  $n_{\rm max} \,{=}\, 4\, n_{\rm peak}$.

\subsection{Parameter Estimation}\label{sec:ParamEstimation}

Once the methodology to derive the SNR from an eccentric MBHB has been framed, the natural subsequent step is determining how well the system parameters can be measured. In the case of high SNR, they can be quickly estimated through the \textit{Fisher Information Matrix} formalism. Specifically, the GW signal we are considering is characterized by 9 free parameters (see their definition in Section~\ref{sec:GWSignal}):
\begin{equation} \label{Eq:free_par}
    \vec{\lambda}=(\zeta,f_k,e,i,\psi,l_0,\gamma,\phi,\theta).
\end{equation}
To reconstruct the most probable source parameters, $\vec{\lambda}$, given a set of data, $\vec{d}$, it is possible to work within the Bayesian framework and derive the posterior probability density function $p(\vec{\lambda}|\vec{d})$:
\begin{equation}
    p(\vec{d}|\vec{\lambda}) \,{\propto}\, p(\vec{\lambda})p(\vec{d}|\vec{\lambda}),
\end{equation}
where $p(\vec{\lambda}| \vec{d})$ is the likelihood function and  $p(\vec{\lambda})$ is the prior probability density of $\vec{\lambda}$. If we assume that near the maximum likelihood estimated value, $\hat{\lambda}_i$, the prior probability density is flat, the posterior distribution $p(\vec{\lambda}| \vec{d})$ will be proportional to the likelihood and can be approximated as a multi-variate Gaussian distribution:
\begin{equation}
  p(\vec{\lambda}| \vec{d})   \,{\propto}\, \mathrm{exp} \left[-\frac{1}{2}\Gamma_{ij}\Delta\lambda_i \Delta\lambda_j\right],
\end{equation}
where the indexes $i$ and $j$ run over all the components of the source parameter vector $\vec{\lambda}$ (in our case from 1 to 9), and $\Delta \lambda_i \,{=}\,  \hat{\lambda}_i \,{-}\,  \lambda_i$ ($\Delta \lambda_j \,{=}\, \hat{\lambda}_j \,{-}\,  \lambda_j$) are the differences between the 'true' source parameters ($\vec{\lambda}$) and their most probable estimated values ($\hat{\lambda}$). Finally, $\Gamma_{ij}$ is the Fisher information matrix, and its inverse provides a lower limit to the error covariance of unbiased estimators\footnote{For further details about the Fisher Information Matrix, we refer the reader to \cite{Husa2009} and \cite{SesanaVecchio2010}.}. In the PTA case, the Fisher matrix is computed as:
\begin{equation} \label{eq:FisherMatrixPTA}
    \Gamma_{ij}  = 4 \int_0^{\infty} df \, \dfrac{\partial_i s(f) \partial_j s(f)}{S_k(f)},
\end{equation}
where $\partial_i$ and $\partial_j$ are the partial derivatives of time residual in the frequency domain, $s(f)$, with respect to the $\lambda_i$ and $\lambda_j$ parameters, respectively.
As for the SNR integral, the scalar product is defined in the frequency domain. However, we can apply the approximate identity given by Eq.~\eqref{eq:innerproduct_f_to_t} to write:
%However, the CGW signal emitted by an eccentric MBHB does not have any analytical form in the residuals nor for its derivative. To overcome this issue we can apply the same method that was used in Section ~\ref{sec:SNR} and write the Fisher Information Matrix as:
\begin{equation}
    \Gamma_{ij} \,{\simeq}\,  \sum_{n=1}^{n_{max}} \dfrac{2}{S_k(f_n)} \int_{t_0}^{t}  dt' \, \partial_i s_n(t') \partial_j s_n(t'),
\end{equation}
in which the partial derivatives are calculated numerically through:
\begin{equation}
    \partial_i s_n(t) \,{=}\, \left[ \frac{   s_n(t,\lambda_i + \delta \lambda_i/2) - s_n(t,\lambda_i - \delta \lambda_i/2)   }{\delta \lambda_i}\right],
\end{equation}
where the time step is set to be equal to $\delta \lambda_i\,{=}\,10^{-5}\lambda_i$. We note that when calculating the SNR and the Fisher information matrix we always assume to know all the parameters that fully specify the residuals $s(t)$.\\

By assuming independent data streams for each pulsar in the array, the Fisher information matrix obtained from the full PTA, $\left( \Gamma_{ij} \right)_{T}$, is simply given by the sum of the single Fisher information matrices derived for each pulsar, $\left( \Gamma_{ij} \right)_k$:
\begin{equation}
    \left( \Gamma_{ij} \right)_{\rm tot} \,{=}\,  \sum_{k\,{=}\,1}^{N_{\rm Pulsars}} \left( \Gamma_{ij} \right)_k.
\end{equation}

We stress that the covariance matrix is simply the inverse of the Fisher information matrix ($\Gamma^{-1}$), thus the elements on the diagonal represent the variances of the parameters ($\sigma_{ii}^2\,{=}\,\Gamma^{-1}_{ii}$), while the off-diagonal terms correspond to the correlation coefficients between parameters ($\sigma_{ij}^2\,{=}\,\Gamma^{-1}_{ij}/\sqrt{\sigma_{i}^2\sigma_{j}^2}$).

\subsection{Characterising the noise} \label{sec:Noise}
The next fundamental ingredient in our computation is the noise description in PTA experiments. In particular, the pulsar NPSD can be broken down in two separate temrs:
\begin{equation} \label{eq:NPSD}
    S_k(f) \,{=}\, S_h(f) +S_p(f).
\end{equation}

The term $S_h(f)$ describes the red noise contributed at each given frequency by the sGWB generated by the incoherent superposition of all the CGWs emitted by the cosmic population of adiabatically MBHBs \citep{Rosado2015}: 
\begin{equation} \label{eq::NPSD_blackhole}
S_h(f) \,{=}\,  \dfrac{h_c^2(f)}{12\pi^2f^3},%  = \dfrac{\sum_{j=1}^{N_{Bin}} h_{j}^2(f)}{12\pi^2f^3}   %S_h(f) = \dfrac{h_c^2(f)}{f} \dfrac{1}{12\pi^2f^2},  
\end{equation}
For a real PTA, the noise is estimated at each resolution frequency bin of the array. In fact if we assume an observation time $T$, the PTA is sensitive to an array of frequency bins $\Delta{f_i}=[i/T,(i+1)/T]$, with $i=1,...,N$. If we now identify each frequency bin $\Delta{f_i}$ with its central frequency $f_i$, then we can associate to each frequency resolution element the characteristic strain produced by all the MBHBs emitting in that element as:
\begin{equation}\label{eq:hc_disc}
    h_c^2(f_i) \,{=}\, \sum_{j=1}^{N_S} h_{c,j}^2(f)\delta(\Delta{f_i}\,{-}\,f).
\end{equation}
where the sum is over all sources, $N_S$, and $\delta(\Delta{f_i}\,{-}\,f)$ is a generalized delta function that assumes the value $1$ when $f\in \Delta{f_i}$, and $0$ otherwise, thus selecting only MBHBs emitting within the considered bin. $h_{c,j}^2(f)$ is the squared characteristic strain of the  $j{-}\rm th$ source.
%Here $j$ is an index running over all the sources in our MBHB population, $N_{Bin}$, and $h_{j}^2(f)$ is the characteristic strain emitted by the $j{-}\rm th$ MBHB at an observed frequency, $f$. 
Since we consider eccentric MBHBs, $h_{c,j}^2(f)$ is the sum of the strain emitted at all the harmonics $nf_k$, among which one has to select only those that lie within the frequency bin $\Delta{f_i}$. Eq.~\eqref{eq:hc_disc} thus generalizes to
\begin{equation}
    h_c^2(f_i) \,{=}\, \sum_{j=1}^{N_S} \sum_{n=1}^{n_{\rm max}}  h_{c,n,j}^2(nf_k) \, \delta(\Delta{f_i}\,{-}\,nf_k).% \, \delta(f-f_{n}),
\end{equation}
%being the final value of $h_j^2(f)$ the sum of only the harmonics whose frequency, $nf_k$, fall within the frequency bin, $f$. 
For each of the $N_S$ binaries the value of $h^2_{c,n}$ is given by: 
\begin{equation}\label{eq:Strain_harminic_n}
    h^2_{c,n} = \frac{\left(\mathcal{A}^2 + \mathcal{B}^2\right)}{2} \dfrac{( G\mathcal{M}_z)^{10/3}}{ c^8 D_L^2 }(2\pi f_k/n)^{4/3} \frac{g(n,e)}{(n/2)^2} \, \dfrac{nf_k}{\Delta f},
\end{equation}
where $\mathcal{A}\,{=}\, 1+ \cos{(i)}^2$, $\mathcal{B}\,{=}-2\cos{(i)}\,$ and
%\begin{equation} \label{SGWB_single}
%       h_{c,n}^2 (f)=  4 \dfrac{32}{5}  \dfrac{(G\mathcal{M}_z)^{10/3}}{n^2 c^8 D_L^2 }(2\pi f_k)^{4/3}g(n,e) \, \dfrac{nf_k}{\Delta f},
%\end{equation}
$\Delta f=1/T$ is the frequency bin width. The value of $g(n,e)$ is computed according to: 
\begin{equation}
        g(n,e) \,{=}\, \dfrac{n^4}{32} \left( B_n^2 + (1-e^2)A^2_n + \dfrac{4}{3n^2}J_n(ne)^2   \right),
\end{equation}
where $A_n$ and $B_n$ are:
\begin{equation}
\begin{aligned}
        B_n & \,{=}\, J_{n-2}(ne) \,{-}\, 2eJ_{n-1}(ne) \,{+}\, \dfrac{2}{n} J_n(ne) \,{+}\, 2eJ_{n+1}(ne) \,{-}\, J_{n+2}(ne), \\
        A_n &\,{=}\, J_{n-2}(ne) \,{-}\, 2J_n(ne) \,{+}\, J_{n+2}.
\end{aligned}
\end{equation}
We stress that when evaluating the detectability of a given MBHB we will not take into account the contribution of its $h_{c,j}^2(f)$ when computing value of $S_h(f)$.\\
%In this way, the total gravitational strain at each observed frequency bin produced by a cosmological population of eccentric MBHB will be:
%\begin{equation}
%    h_c^2(f) = \sum_i h_{c,n}(f)_i,
%\end{equation}
%where $i$ is the index of all the harmonics whose contribution to the total strain falls in the same frequency bin. \davcoment{We stress that when evaluating the detectability of a given MBHB we will not take into account the contribution of its $h_c^2(f)$.}\\ %, the effect of the eccentricity is to push the signal towards higher frequencies as an outcome the slope of the GWB is flattened at low frequencies, as can be seen in Figure, where the GWB for three different SMBHB populations characterized by a different eccentricity distribution is displayed.

%\begin{figure}[h!] 
%    \centering
%    \hspace{0.005\textwidth}
%    \includegraphics[width=0.5\textwidth]{Figures/GWB_3.png}
%    \caption{Caption for the figures.}
%    \label{Fig::GWB}
%\end{figure}
The term $S_p(f)$ in the NSPD encodes all sources of noise unrelated to the sGWB, which are related to the telescope sensitivity, intrinsic noise in the pulsar emission mechanism, pulse propagation effects and so on. PTA collaborations parameterize the pulsar noise as a combination of three different terms:
\begin{equation} \label{Eq::tot_NPSD}
    S_p(f) \,{=}\, S_{w} \,{+}\, S_{\rm DM}(f) \,{+}\, S_{\rm red}(f).
\end{equation}
$S_{w}$ accounts for processes that generate a white stochastic error in the measurement of a pulsar arrival time. These include pulse jitter, changes in the pulse profile with time, or instrumental artefacts. Such processes are uncorrelated in time and the resulting noise is modelled as $S_{w} \,{=}\, 2\Delta t_{\rm cad} \sigma^2_{w}$, where it is commonly assumed that the pulse irregularity is a random Gaussian process described by the root mean square value $\sigma_{w}$. $\Delta t_{\rm cad}$ is the time elapsed between two consecutive observations of the same pulsar, i.e. the observation cadence. $S_{\rm red}(f)$ and $S_{\rm DM}(f)$ describe the achromatic and chromatic red noise contributions, respectively. 
%red noises and are so-called (achromatic) \textit{red noise} and \textit{dispersion measure} (DM chromatic) \textit{noise}, respectively. 
While the former is the result of the pulsar intrinsic noise, the latter is the result of spatial variations in the interstellar electron content along the line of sight between the observer and the pulsar. These two red noises are usually modeled as a stationary stochastic process, described as a power law and fully characterized by an amplitude and a spectral index. 

%In the following and unless otherwise stated, we include only $S_{w}$ in our calculations; the validity of this assumption is discussed in Appendix \ref{appendix:RedNoiseEffect}.

\section{Supermassive black hole binary populations and Pulsar Timing Arrays} \label{sec:pop and PTA}

%In this section, we present the population of MBHBs and Pulsar Timing Arrays that will be used.

\subsection{The population of binaries} \label{sec:SMBHB_population}

In this section, we briefly present the procedure used to generate the different populations of eccentric MBHBs that will be used throughout this paper. For further details, we refer the reader to \cite{2010ApJ...719..851S, Sesana2013} and \cite{InterpretationPaperEPTA2023}.\\

To study the detectability of single MBHBs it is required to characterise their cosmological population as a whole. The sGWB spectrum generated by such a population can be calculated as the integrated emission of all the CGW signals emitted by individual binaries. Thus, the inclination and polarization average\footnote{The sky and polarization average implies that $\left(\mathcal{A}^2 + \mathcal{B}^2\right) = 64/5$} characteristic strain of the sGWB can be expressed as:
\begin{equation} \label{eq:SGWB_Population}
   \begin{aligned}
h_c^2(f)=\int_0^{\infty} \mathrm{d} z \int_0^{\infty} \mathrm{d} m_1 \int_0^1 \mathrm{~d} q \frac{\mathrm{d}^5 N}{\mathrm{~d} z \mathrm{~d} m_1 \mathrm{~d} q \mathrm{~d}e \mathrm{~d} \ln f_{\mathrm{k}, r}} \times \\
\left.\frac{32}{5}\frac{(G\mathcal{M}_z)^{10/3}}{c^8 D_L^2 (1+z)^{4/3}} (2\pi f_{\mathrm{k}, r})^{4/3} \sum_{n=1}^{\infty} \frac{g\left(n, e\right)}{(n / 2)^2}\right|_{f_{\mathrm{k}, r}=f(1+z) / n},
\end{aligned}
\end{equation}
where $\mathrm{d}^5N/(\mathrm{~d} m_{1} \mathrm{~d}q \mathrm{~d}z \mathrm{~d}e \mathrm{~d}t_r)$ is the comoving number of binaries emitting in a given logarithmic frequency interval, $\mathrm{~d} \ln f_{\mathrm{k}, r}$, and primary mass, mass ratio, eccentricity and redshift in the range [$m_{1}$,$m_{1} + \delta m_{1}$], [$q, q + \delta q$], [$e$, $e + \delta e$] and [$z$, $z+\delta z$], respectively. In particular, this quantity can be re-written as:
\begin{equation} \label{eq:Comoving_Merger}
\begin{aligned}
    \frac{\mathrm{d}^5 N}{\mathrm{~d} z \mathrm{~d} m_1 \mathrm{~d} q \mathrm{~d}e \mathrm{~d} \ln f_{\mathrm{k}, r}}= \frac{d^3n}{\mathrm{~d} z \mathrm{~d} m_1 \mathrm{~d} q} \left( \frac{1}{f_{\mathrm{k},r}}\frac{df_{\mathrm{k},r}}{dt_r}  \right)^{-1} \left[ \frac{\mathrm{~d} z}{\mathrm{~d} t_r} \frac{\mathrm{~d} V}{\mathrm{~d} z}\right] \\
     = \frac{d^3n}{\mathrm{~d} z \mathrm{~d} m_1 \mathrm{~d} q } \left( \frac{1}{f_{\mathrm{k},r}}\frac{df_{\mathrm{k},r}}{dt_r}  \right)^{-1}  \,  \, \frac{4\pi c D_L^2}{(1+z)},
    \end{aligned}
\end{equation}
where $n\,{=}\,dN/dV$, $d^3n /(\mathrm{~d} z \mathrm{~d} m_1 \mathrm{~d} q)$ is the differential merger rate comoving density of MBHBs and $f_{\mathrm{k},r} (\mathrm{~d} t_r / \mathrm{~d} f_{\mathrm{k},r})$ represents the binary evolution timescale, which implicitly takes into account the variation of the hardening rate with the binary eccentricity (i.e. at fixed orbital frequency, eccentric binaries evolve faster). Following \cite{Sesana2013}, the merger rate of MBHs can be expressed in terms of the galaxy merger rate, ($\mathrm{d}^3n_{\rm G} /(\mathrm{d} z \, \mathrm{d} M_{*} \, \mathrm{d}q_{*})$, as:
\begin{equation}\label{eq:Galaxy_and_MBH_Merger_ratio}
\begin{split}
   \frac{d^3n}{\mathrm{~d} z \, \mathrm{~d} m_{1} \, \mathrm{~d} q} & = \frac{\mathrm{~d}^3n_{\rm G}}{\mathrm{~d} z \, \mathrm{~d} M_{*} \, \mathrm{~d}q_{*}}  \frac{\mathrm{~d} M_*}{\mathrm{~d} m_{1}}  \frac{\mathrm{~d} q_*}{d q} =  \\ &
   {=} \, \left[ \frac{\phi(M_*,z)}{M_* \ln 10} \frac{\mathcal{F}(z,M_*,q_{*})}{\tau(z,M_*,q_{*})} \frac{dt}{dz} \right] \frac{\mathrm{~d} M_*}{\mathrm{~d} m_{1}}  \frac{\mathrm{~d} q_*}{d q},
\end{split}
\end{equation}
where $\phi(M_*,z)$ is the galaxy stellar mass function and $\mathcal{F}(z,M_*,q_*)$ the differential fraction of galaxies with mass $M_*$ at a given redshift paired with a satellite galaxy of mass in the interval $[q_* M_*, (q_*\,{+}\, dq_*) M_{*}]$\footnote{Specifically, $\mathcal{F}(z,M_*,q_*)$ was computed by setting $ \mathcal{F}(z,M_*,q_*) \,{=}\, -f_0(1+z)^{\gamma}/(q_*\ln q_m)$, being $f_0$ and $\gamma$ free parameters inferred from observational studies and $q_m$ the minimum mass ratio selected in counting pairs.}. The value $\tau(z,M_*,q)$ is deduced from N-body simulations and corresponds to the typical merger timescale for a galaxy pair with a given mass, redshift and mass ratio. The term $(d M_*/ d m_1) (dq_*/dq)$ associates an MBH to each galaxy in the pair by using the MBH galaxy bulge mass scaling relation: 
\begin{equation}\label{eq:ScalingRelation_SMBHs}
    \log_{10}(M_{\rm BH}) = \alpha + \beta \log_{10}(M_{\rm Bulge}) + \mathcal{E}
\end{equation}
where $\mathcal{E}$ represents an intrinsic scatter, generally around $0.3-0.5$ dex \citep{Sesana2016}, and $\alpha$ and $\beta$ define the zero point and logarithmic slope of the relation, respectively. To transform the total stellar mass into bulge mass, the relation $M_{*}\,{=}\, f_{\rm b} \,M_{\rm Bulge}$ described in \cite{Sesana2016} is assumed.\\

%\as{I simplified this section since it was too dense. In particular, I removed the accretion part since that's true mostly for David's models, whereas in my model a number of accretion scenarios can occur. In any case, this is kind of irrelevant details here.}
%For elliptical galaxies with $M_{*} \,{>}\, 10^{11} \, M_{\odot}$ ($M_{*} \,{<}\, 10^{11} \, M_{\odot}$) $f_{\rm b}\,{=}\, 1$ $(0.5)$, while for late-type systems, a random $f_{\rm b}$ in the range $0.1\,{-}\,0.3$ was taken. 
%It is worth noticing that the final MBH mass assigned to the remnant galaxy ($M_{\rm BH}$) is larger than the masses of the two progenitors ($M_{\rm BH} \,{<}\, m_1 \,{+}\, m_2$). This is due to the fact that MBHs can grow during galactic encounters through gas accretion. To assign the extra mass to one of the two MBH progenitors ($M_{\rm acc} \,{=}\,M_{\rm BH} \,{-}\, m_1 \,{-}\, m_2$), we followed the recent results from hydrodynamical simulations stating that during galaxy interactions the secondary galaxy suffers large perturbations during the pericenter passages around the central one \citep[e.g][]{Capelo2015}. In these circumstances, the black hole of the secondary galaxy experiences accretion enhancements. Thus, the mass of the secondary MBHB drawn from Eq.~\eqref{eq:ScalingRelation_SMBHs} is raised by an amount $M_{\rm acc}$ \citep{Sesana2009}.\\ 
%\begin{equation}
%     \alpha = \frac{M_{BH} - m_2 - m_1}{m_2}
% \end{equation}

Finally, the hardening of the binary in Eq.~\eqref{eq:Comoving_Merger} is determined by using the stellar models of \cite{Sesana2010}:
\begin{equation}\label{eq:frequency_Evolution}
\begin{split}
\frac{df_{k,r}}{dt_r} & \,{=}\, \left( \frac{df_{k,r}}{dt_r} \right)_{*}  \, {+} \, \left( \frac{df_{k,r}}{dt_r}  \right)_{GW} \, {=} \, \\ &
\,{=}\, \frac{3G^{4/3} (m_{1} + m_{2})^{1/3} H \rho_i }{2(2\pi)^{2/3} \sigma_i}f_{k,r}^{1/3} \, {+} \, \frac{96(G \mathcal{M})^{5/3}}{5c^5} (2\pi)^{8/3} f_{k,r}^{11/3}\mathcal{F}(e),
\end{split}
\end{equation}
and 
\begin{equation}  \label{eq:eccentricity_Evolution}
\begin{split}
& \frac{de}{dt_r}  \,{=}\, \left( \frac{de}{dt_r} \right)_{*}  \, {+} \, \left( \frac{de}{dt_r}  \right)_{GW} \, {=} \,  \\ & 
\,{=} \, \frac{G^{4/3}(m_1 + m_2)^{1/3} \rho_i H K }{(2\pi)^{2/3} \sigma_i}  f_{k,r}^{-2/3} \, {-} \, \frac{(G \mathcal{M})^{5/3}}{15c^5} (2\pi f_{k,r})^{8/3}  \mathcal{G}(e),
\end{split}
\end{equation}
where 
\begin{equation}
\begin{split}
  & \mathcal{F}(e)   = \frac{1+(73/24)e^2 + (37/96)e^4}{(1-e)^{7/2}}, \\ & 
  \mathcal{G}(e) = \frac{304e + 121e^3}{(1-e^2)^{5/2}},
\end{split}
\end{equation}
and $\sigma_i$ and $\rho_i$ are the velocity dispersion and stellar density at the binary influence radius. $H$ and $K$ represent the hardening rate and the eccentricity growth rate, calibrated against numerical three-body scattering experiments \citep{Sesana2006}. 

\subsubsection{Generating MBHB populations consistent with PTA measurements}

As described above, the cosmological coalescence rate of MBHBs depends on different assumptions about the galaxy merger rate and correlations between MBHBs and their hosts. In particular, the library of models presented in \cite{Sesana2013,Rosado2015,Sesana2016} combines a number of prescriptions from the literature which we summarize here: 

\begin{enumerate}
    \item \textit{Galaxy stellar mass function}. Five different observational results are taken from the literature \citep{Borch2006, Drory2009, Ilbert2010, Muzzin2013, Tomczak2014} and matched with the local mass function \citep{Bell2003}. For each of these functions, upper and lower limits were added to account for the errors given by the authors best-fit parameters. On top of this, an additional $0.1\, \rm dex$ systematic error was included to consider the uncertainties in the stellar masses determination. For all the mass functions, we separate between early/late-type galaxies and the analysis was restricted to $z\,{<}\,1.3$ and $M_* \,{>}\,10^{10}\,M_{\odot}$, since we expect that these systems contribute the most to the sGWB signal \citep[see e.g][]{Sesana2009,Kelley2018,IzquierdoVillalba2023} \\ 
    
    \item \textit{Differential fraction of paired galaxies}. The observational results of \cite{Bundy2009}, \cite{deRavel2009}, \cite{Lopez-Sanjuan2012} and \cite{Xu2012} were used when accounting for the evolution of the galaxy pair fraction. \\
    
    \item \textit{Merger timescale for a galaxy pairs}. We follow the fits done from the N-body and hydrodynamical simulations of \cite{Kitzbichler2008} and \cite{Lotz2010}. \\
    
    \item \textit{Galaxy-MBH scaling relation}. The masses assigned to each merging galaxy pair were drawn from several observational relations. However, given the high normalization of the observed PTA signal, we only considered relations presented by  \cite{Kormendy2013,McConnell2013} and \cite{Graham2013}\\
\end{enumerate}

\noindent To save computation time, we perform an ad-hoc down-selection of the models, and limit our investigation to 108 combinations of the above prescriptions producing a distribution of sGWB amplitudes consistent with the measured PTA signal, as per Figure 2 of \citep{InterpretationPaperEPTA2023}.\\ 

As for the environmental coupling and eccentricity evolution, we adopt the following prescriptions:
    
\begin{enumerate}
    \item \textit{Stellar density profile}. Following \cite{Sesana2010}, the stellar density profile is assumed to be a broken power law following an isothermal sphere outside the influence radius, $r_{i} \,{=}\, 1.2\,{\rm pc}\, (M/10^6 M_{\odot})^{0.5}$, and a profile
    \begin{equation} \label{density}
    \rho \,{=}\, C \rho_i\left( \frac{r}{r_{i}} \right)^{-1.5}
    \end{equation}
    at $r<r_i$. Here, $\rho_i \,{=}\, \sigma^2 / (2\pi G r^2_i)$ and $\sigma$ is determined from  the \cite{Tremaine2002} scaling relation \citep[see][for further details]{Sesana2010}. $C$ is a normalization factor of the stellar density profile and is assumed to take three different values ($0.1$, $1$ and $10$), to investigate the effect of changing the typical density of the environment.\\

    \item \textit{Initial eccentricity}. During the tracking of the hardening evolution, all the binaries are assumed to start with an initial eccentricity $e_0$ at binary formation. Throughout the paper, we consider $10$ initial values of $e_0\,{=}\, 0, 0.1, 0.2, 0.3, 0.4, 0.5, 0.6, 0.7, 0.8, 0.9$.
\end{enumerate}

\noindent Using the 108 population model, 10 eccentricity values and 3 environment normalizations defined above, we generate 3240 numerical distributions of MBHBs using Eq.~\eqref{eq:Comoving_Merger}, and for each distribution, we perform 10 Monte Carlo sampling for a gran total of 32400 MBHB populations. Each population consists of a list of $\approx10^5$ binaries characterized by their chirp mass, redshift, orbital frequency and eccentricity. Due to the computational cost required to compute the fisher information matrix, the latter has been calculated only for a subsample of 3240 populations (10\% of the total).

%\noindent Taking into account the stellar mass function, galaxy pairs, galaxy merger time, and SMBH-galaxy correlation it was possible to generate 108 populations of merging MBHs whose sGWB amplitude at 1$yr^{-1}$ ranges from $10^{-15}$ to $10^{-14.5}$, in agreement with recent PTA results \citep{Antoniadis2023,Agazie2023,Reardon2023,Xu2023}. Afterward, 100 Monte Carlo samplings were generated from these merging systems and evolved according to Eq.~\eqref{eq:frequency_Evolution} and Eq.~\eqref{eq:eccentricity_Evolution} taking as initial conditions the $10$ different values of $e_0$ and the $3$ normalization values of $C$. Thus, a total number of $108\,{\times}\,10\,{\times}\,3\,{\times}\,10\,{=}\,32400$ eccentric populations were generated for this work. However, due to the computational cost required to compute the fisher information matrix, the latter has been calculated only for a subsample of these populations specifically 3240.% We stress that we will group the results according to the values of $e_0$.

\subsection{The array of pulsars} \label{sec:Pulsar_Arrays}
We explore the feasibility of detecting CGW signals using two different pulsar timing arrays: MeerKAT ($N_{\rm Pulsars}\,{=}\,78$) and SKA ($N_{\rm Pulsars}\,{=}\,200$). While pulsar monitoring at MeerKAT has been ongoing for 4.5 years, it will be superseded by the SKA Mid array by 2027. Given these constraints, we choose 10 years for MeerKAT while the 30-year time span of SKA follows projections commonly used in the literature.\\

\begin{figure}
    \centering
    \hspace{0.005\textwidth}
    \includegraphics[width=1.0\columnwidth]{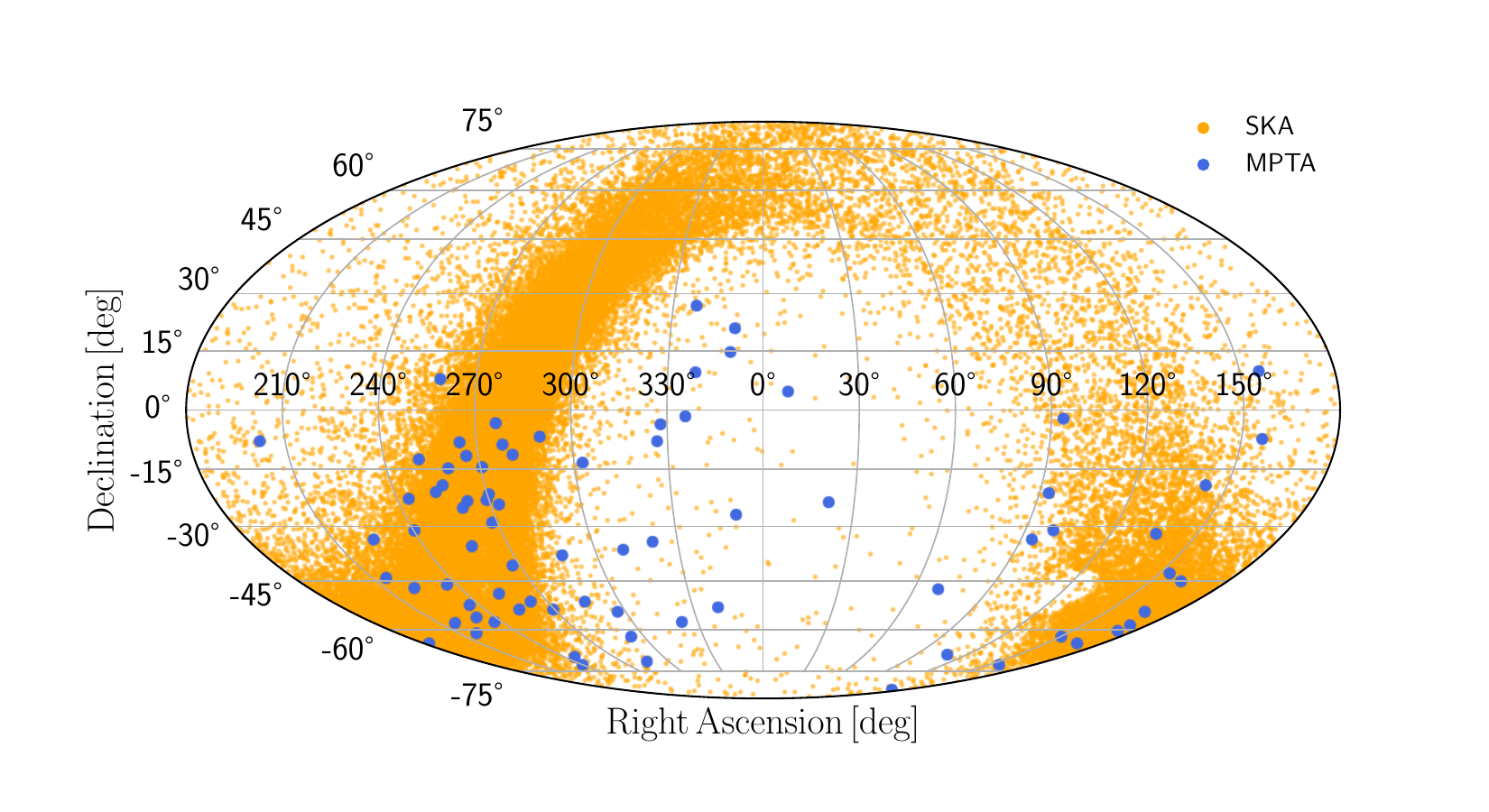}
    \caption{Sky position of the pulsars included in the PTA experiments used in this work. The pulsars from MPTA are displayed in blue whereas the ones of SKA (full \texttt{PsrPopPY} population) are depicted in orange.}
    \label{fig:Pulsars_Sky_Location}
\end{figure}

\noindent i) \textbf{MeerKAT} is a 64-antenna radio interferometer telescope located in South Africa. The regular monitoring of millisecond pulsar timing by MeerKAT is the basis of the MPTA. %\textit{MeerKAT Pulsar Timing Array} (MPTA, \citealt{Bailes2020}). 
Recently, it was released the initial 2.5-years MPTA data \citep{Miles2023}. While the current data includes 88 pulsars, the release only contains the 78 pulsars that have at least 30 observations over this observing span, with a typical cadence of 14 days. The upper panel of Fig.~\ref{fig:Pulsars_Sky_Location} shows the position of those pulsars in the sky. Table A1 of \cite{Miles2023} also reported the noise properties of each of the 78 pulsars, accounting for white-noise terms, frequency-dependent DM variations, and an achromatic red-noise process (see Section~\ref{sec:Noise}). In this work, we will use a 10-year MPTA-like system, featuring the same set of pulsars (number, sky position, and noise model) as the one presented in \cite{Miles2023}. \\

\begin{figure}
    \centering
    \includegraphics[width=1\columnwidth]{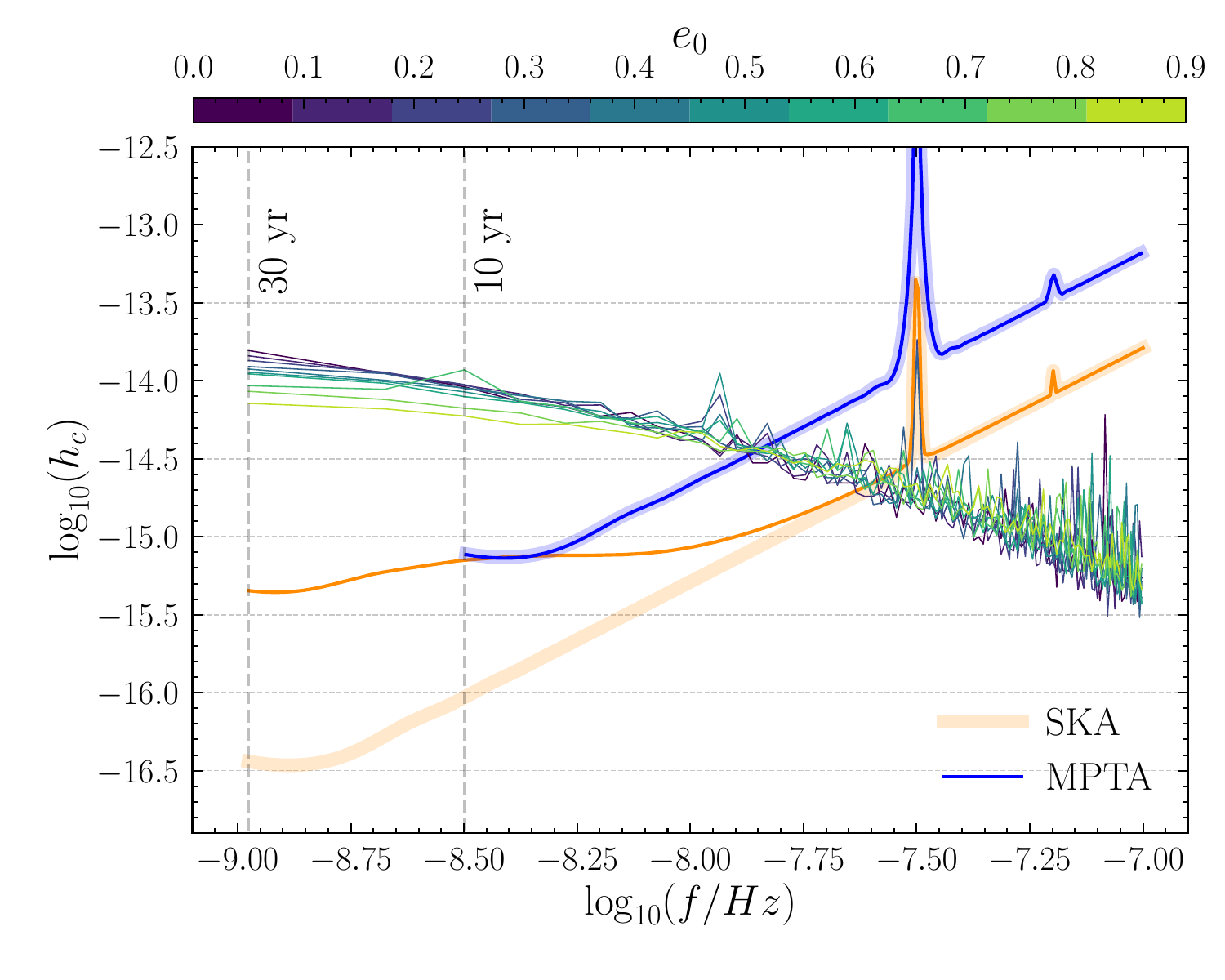}
    \caption{Sensitivity curves of 10-year MPTA (blue) and 30-year SKA (orange) computed from \texttt{HASASIA}. The vertical dashed lines correspond to the frequency associated with the observing time span of MPTA and SKA (highlighted with vertical dashed lines). For completeness, the square mean of sGWB produced by our MBHB populations at different $e_0$ models are represented in different colors. Pale (dark) blue and orange lines correspond to the MPTA and SKA sensitivity curves when accounting for pulsar white noise only (white plus red noise).}
    \label{fig:Sensitivuty_Curve}
\end{figure}

\noindent ii) \textbf{Square Kilometer Array Mid telescope} (SKA, \citealt{Dewdney2009}) planned to be operative in 2027, will be a large radio interferometer telescope whose sensitivity and survey speed will be an order of magnitude greater than any current radio telescope. For this work, we simulate a 30-year SKA PTA with 200 pulsars featuring a white noise of $\sigma_w\,{=}\,100\, ns$ and an observing cadence of 14 days. To picture a more realistic scenario we also include red noise to the total noise power spectral density in Eq. \eqref{Eq::tot_NPSD}, parameterized as a power law \citep[see e.g][]{Lentati2015} of the form 
\begin{equation}
S_{\rm red}(f) = \dfrac{A_{\rm red}^2}{12 \pi^2} \bigg( \dfrac{f}{f_{\rm yr}}\bigg)^{-\gamma_{\rm red}} \rm yr^{3},
\end{equation}
where $A_{\rm red}$ the amplitude at one year and $\gamma_{\rm red}$ the spectral index. Red noise properties are drawn to be consistent with those measured in the EPTA DR2Full using the following procedure. We fit a linear ${\rm log}\,A_{\rm red}-\gamma$ relation to the measured red noises in Table 4 of \cite{Custom_noise_EPTA}. We then assign $A_{\rm red}$ and $\gamma$ parameters consistent with this relation to 30\% of the pulsars in the SKA array, drawing $A_{\rm red}$ randomly from a uniform log-distribution in the range to $-15<{\rm log}A_{\rm red}<-14$, for which the corresponding $\gamma$ is $>3$. In this way, we mimic in the SKA array the fraction and properties of EPTA DR2Full pulsars with a robust red noise contribution. Note that while the remaining 70\% of the pulsars will likely display some lower level of red noise, this is unlikely to affect the properties of the detected CGWs. In fact, for those pulsars the main stochastic red noise component is going to be the sGWB itself, which is already included in our calculation. 

%We consider only DR2Full pulsars with red noise detection with $\gamma>3.$ and fit a linear ${\rm log}A-\gamma$ relation. We then assign to 30\% of the selected pulsars in the SKA array a value of $A_{\rm red}$ randomly from a uniform distribution in the range to $10^{-15}<A_{\rm red}<10^{-14}$ and $\gamma_{\rm red}$ is then assigned according to a log-linear relation between $A_{\rm red}$ and $\gamma_{\rm red}$ obtained by fitting the red noise properties displayed in \cite{Custom_noise_EPTA}. 
%assign red noise to 30\%  of the pulsars, consistent with the fraction of pulsars that show evidence of red noise in the EPTA DR2full \citep{Custom_noise_EPTA}. In particular, the contribution of red noise
%\riccoment{ We set the maximum amplitude $A_{\rm red}=10^{-14}$ since we aim to select only the pulsar that shows clear evidence of red noise. While pulsars that show higher amplitude also display a flatter slope, it could be due to the white noise being reabsorbed in the red noise model. }. 

\noindent We then employ the pulsar population synthesis code \texttt{PsrPopPY}\footnote{\hyperlink{https://github.com/samb8s/PsrPopPy}{https://github.com/samb8s/PsrPopPy}.} \citep{Bates2014} to draw a realistic distribution of pulsars in the array. \texttt{PsrPopPY} generates and evolves realistic pulsar populations drawn from physically motivated models of stellar evolution and calibrated against observational constraints on pulse periods, luminosities, and spatial distributions. The final population of pulsars ($10^5$) is selected such that they would be observable ($\rm SNR\,{>}\,9$) by a SKA survey with an antenna gain of 140 K/Jy and integration time of 35 minutes. In order to avoid a particularly lucky/unlucky pulsar sky disposition, from this distribution we select a different set of 200 pulsars for each one of the MBHB population presented in Section \ref{sec:SMBHB_population}. The sky distribution of the whole pulsar sample of SKA is presented Fig.~\ref{fig:Pulsars_Sky_Location}. Since \texttt{PsrPopPy} simulates hyper-realistic distributions of pulsars generated using theoretical considerations and observational constraints, the bulk of the generated full population of pulsars will lie close to the Galactic plane. However, the exceptional sensitivity of the SKA would also allow to choose the most isotropic distributions of pulsars in the PTA, maximizing sensitivity to any GW signals searched for by the PTA.\\

Fig.~\ref{fig:Sensitivuty_Curve} presents the sensitivity curve of our SKA PTA and MPTA computed using \texttt{HASASIA} Python package \citep{Hazboun2019}. As expected, SKA PTA features better sensitivity than the MPTA. However, at low frequencies, both of them are limited by the sGWB. To guide the reader, Fig.~\ref{fig:Sensitivuty_Curve} also displays the sensitivity curves of SKA and MeerKAT PTAs when only white noise is considered. As shown in Fig.~\ref{fig:Sensitivuty_Curve}, for MPTA the two sensitivity curves are almost identical since only 4 of the 78 pulsars listed in \cite{Miles2023} have a reported red noise. Conversely, when achromatic red noise is included in the SKA PTA, due to the larger fraction of pulsars affected by it, the red noise slightly hider the array's sensitivity at the lowest frequencies (${<}\,10^{-8} \, \rm Hz$). %\riccoment{I add the SKA sensitivity curve computed with red noise for all the 200 pulsar, computed from an ${\rm log}A-\gamma$ relation shifted down by 0.5dex }% \as{Riccardo, can you try two things with HASASIA? 1-include red noise according to the same prescription to *all* pulsars and 2-add red noise also to the remaining 70\% of the pulsars, but from an ${\rm log}A-\gamma$ relation shifted down by 0.5dex. I'm curious to see how the sensitivity curve appears.}

%To guide the reader, Fig.~\ref{fig:Sensitivuty_Curve} also displays the sensitivity curves of SKA and MeerKAT PTAs when we add a pulsar red noise. For MPTA, the values of $A_{\rm red}$ and $\gamma_{\rm red}$ were taken from the values reported in \cite{Miles2023}\footnote{Note that the MPTA curve is poorly affected given that only 4 of the 78 pulsars listed in \cite{Miles2023} have a reported red noise.}. As shown in Fig.~\ref{fig:Sensitivuty_Curve}, the pulsar red noise slightly hinders the sensitivity of our PTA. Finally, Fig.~\ref{fig:Sensitivuty_Curve} shows the dependence of the sGWB low-frequency bending as a function of the initial eccentricity of the MBHB population.

\subsection{Identifying individually resolvable MBHB} \label{sec:ResolvingBinaries}

\begin{figure}
    \centering
    \includegraphics[width=1\columnwidth]{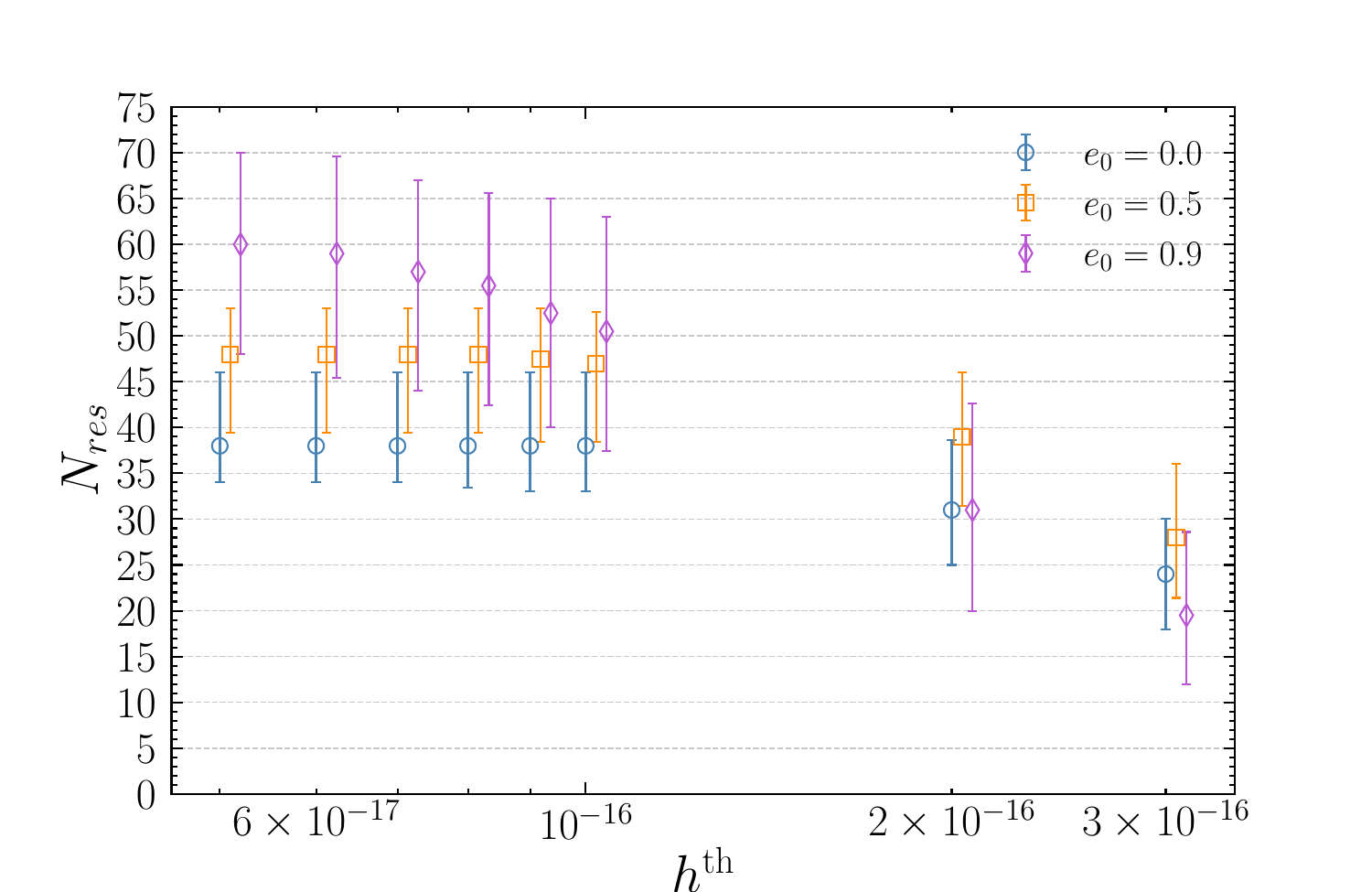}
    \caption{Median number of resolvable sources of 96 randomly selected catalogs ($N_{\rm res}$), with three diffented initial eccentricity,  $e_0=0.0,e_0=0.5,e_0=0.9$,  as a function of the adopted thresholds in the GW signal amplitude ($h^{\rm th}$). Each set of marks and colors represents populations with different $e_0$, as labeled in the figure. The values of $N_{\rm res}$ are computed by using the SKA PTA.}
    \label{fig:h_threshold}
\end{figure}

To extract individually resolvable CGWs, we employ a recursive technique similar to \cite{2021PhRvD.104d3019K} and \cite{2023PhRvD.108j3039P}. We sort the MBHB population by strain amplitude according to the expression in Eq.~\eqref{eq:Strain_harminic_n}, but selecting only the second harmonic ($n\,{=}\,2$). Following this ranking, we calculate the SNR of each source according to Eq.~\eqref{eq:SNR_tot_ecc}, including in the sGWB contribution to the noise the signals produced by all the other MBHBs. Whenever one source exceeds the $\rm SNR\,{>}\,5$ for SKA or $\rm SNR\,{>}\,3$ for MPTA, the source is deemed resolved and its contribution to the sGWB is subtracted. As a consequence, the level of the noise in the pulsar array is lowered as well (see Eq.~\eqref{eq::NPSD_blackhole}), making more feasible the detection of dimmer CGWs that might be otherwise unobservable. We therefore re-evaluate the 
detectability of all the remaining sources by making use of the new (lowered) background. This procedure is repeated until there are no resolvable sources left in the analyzed MBHB population.

\begin{figure*}
    \centering
    \includegraphics[width=1.55\columnwidth]{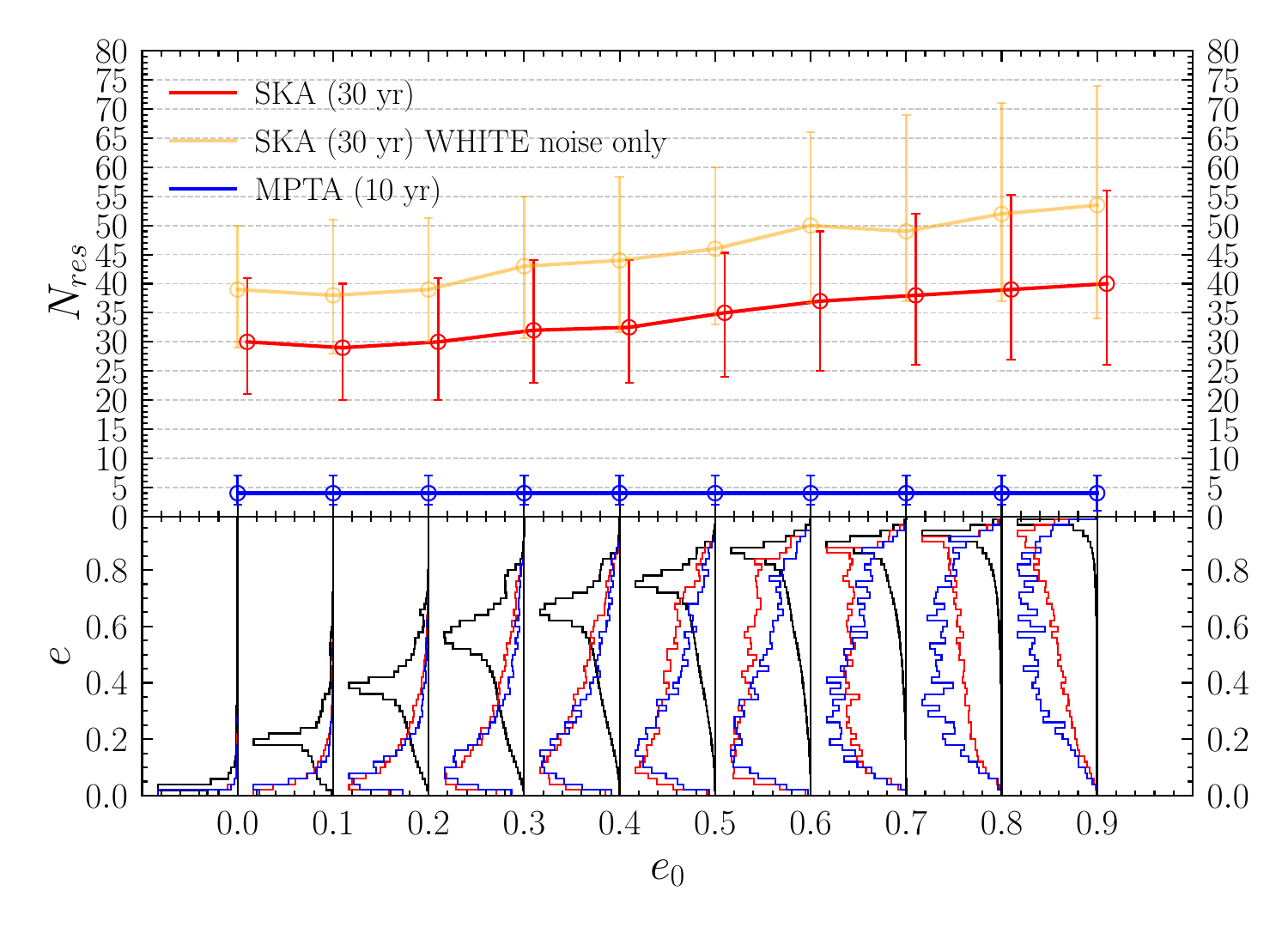}
    \caption{Number of resolvable sources ($N_{\rm res}$) for as a function of $e_0$. Open dots represent median values from all the considered MBHB population models, while error bars represent the 84 and 16 percentile of the distribution. Blue points correspond to the results of 10-year MPTA data with $ \rm SNR\,{>} \, 3$, while the red ones represent the predictions for 30-year SKA data with $\rm SNR\,{>} \, 5$. Orange points represent the  30-year SKA results but when accounting only for the white pulsar noise. The lower panel shows the eccentricity distribution of the whole MBHB population for different $e_0$ models (black) and the eccentricity distribution of the MBHBs detectable as CGW sources ($e_{\rm rs}$) for SKA PTA (orange) and for MPTA (blue). Note that all the distributions are normalized to the same peak value for visualization purposes.}
    \label{fig:Resolvable_Sources_Eccentricity_MeerKAT_SKA}
\end{figure*}

The above recursive procedure must be applied to several thousands of MBHB populations, each including ${\sim}\,10^5$ systems, which becomes extremely time-consuming.
%Computing the SNR as described in The computation of SNR many MBHB populations have been created for this work. These contain a number of MBHBs that can be large enough to hinder a fast and efficient analysis of CGW search. 
To boost the efficiency of our pipeline, we established a criterion that allows us to select only those sources with the largest chance of being resolvable. We established a threshold in the value of $h\,{=}\,2\zeta G^{5/3} (\pi f_{k})^{2/3} / c^{4}$ (hereafter $h^{\rm th}$) below which we do not compute the SNR, deeming the source too dim to be resolved. To determine the exact value of the threshold, we have computed the number of resolvable sources ($N_{\rm res}$) at different $h^{\rm th}$ cuts for 96 randomly selected MBHB catalogs at three different values of $e_0$. We imposed the condition $\rm SNR\,{>}\,5$ for CGW detection and computed the number of resolvable sources using the SKA PTA, because of its larger performance in resolving dim GW sources compared to MPTA. Since, for this analysis, we are interested in the dimmest MBHB that the PTA experiment can resolve, we conservatively consider SKA PTA which features only white noise. %\davcoment{Is this true? I thought Fig.3 was re-done for the new SKA (red+white noise). To me, it is a bit misleading to use only white noise (especially because we defined SKA MPTA with red noise). We should be consistent and do Fi.3 with SKA white+red} \as{I think this is fine. We are just selecting a minimum threshold for resolvable sources. By using white noise only, we are just more conservative because if sources are NOT detectable with white noise only they certainly won't be detectable also with white$+$red noise. I don't think the figure has been redone, but I think it's not necessary}. 
Fig.~\ref{fig:h_threshold} shows the median number of $N_{\rm res}$ as a function of $h^{\rm th}$. As expected, $N_{\rm res}$ increases towards small values of $h^{\rm th}$, but it saturates below a certain threshold. This behavior is seen for all $e_0$ used to start the MBHB evolution. Taking into account Fig.~\ref{fig:h_threshold}, throughout this work we will use the conservative value of $h^{\rm th} \,{=}\,6\,{\times}\,10^{-17}$. We stress that small fluctuations are seen in the $N_{\rm res}$ median below our fiducial threshold. However, they are not statistically significant (${\pm}\,1$ source) and the selected $h^{\rm th}$ provides a good compromise between accuracy and computational efficiency. The recursive SNR evaluation-subtraction procedure is thus performed only on the subset of binaries with $h\,{>}\,h^{\rm th}$, providing a considerable speedup of the calculation.

%Once the value of $h$ is determined for all the MBHB population and only systems with $h\,{>}\,h^{\rm th}$ are selected, we carry out our procedure to resolve and extract sources from the background. To do so, we sort the remaining MBHB population from the largest to the smallest GW strain amplitude according to the expression in Eq.~\eqref{eq:Strain_harminic_n}, but selecting only the second harmonic ($n\,{=}\,2$). Following this ranking, we calculate the SNR of each source according to Eq.~\eqref{eq:SNR_tot_ecc} and whenever one exceeds the $\rm SNR\,{>}\,5$ for SKA or $\rm SNR\,{>}\,3$ for MPTA, the source is deemed resolved and its contribution to the sGWB is subtracted. As a consequence, the level of the noise in the pulsar array is lowered as well (see Eq.~\eqref{eq::NPSD_blackhole}), making more feasible the detection of other CGWs that might be otherwise obscured. Afterward, we re-evaluated the detectability of all the remaining sources by making use of the new (lowered) background. This procedure is repeated until there are no resolvable sources left in the analyzed MBHB population. 

\section{Results} \label{sec:Results}
In this section, we present the main results of our work. The analysis has been performed taking into account different values of $e_0$. This has enabled us to characterize the effect of eccentricity in determining the number of resolvable sources and the accuracy of the parameter estimation from the detected signal. To avoid confusion with the initial eccentricity used in the hardening model, $e_0$, throughout the whole section we will tag the eccentricity of the detected MBHB as $e_{\rm rs}$.

\begin{figure}
    \centering
    \includegraphics[width=1\columnwidth]{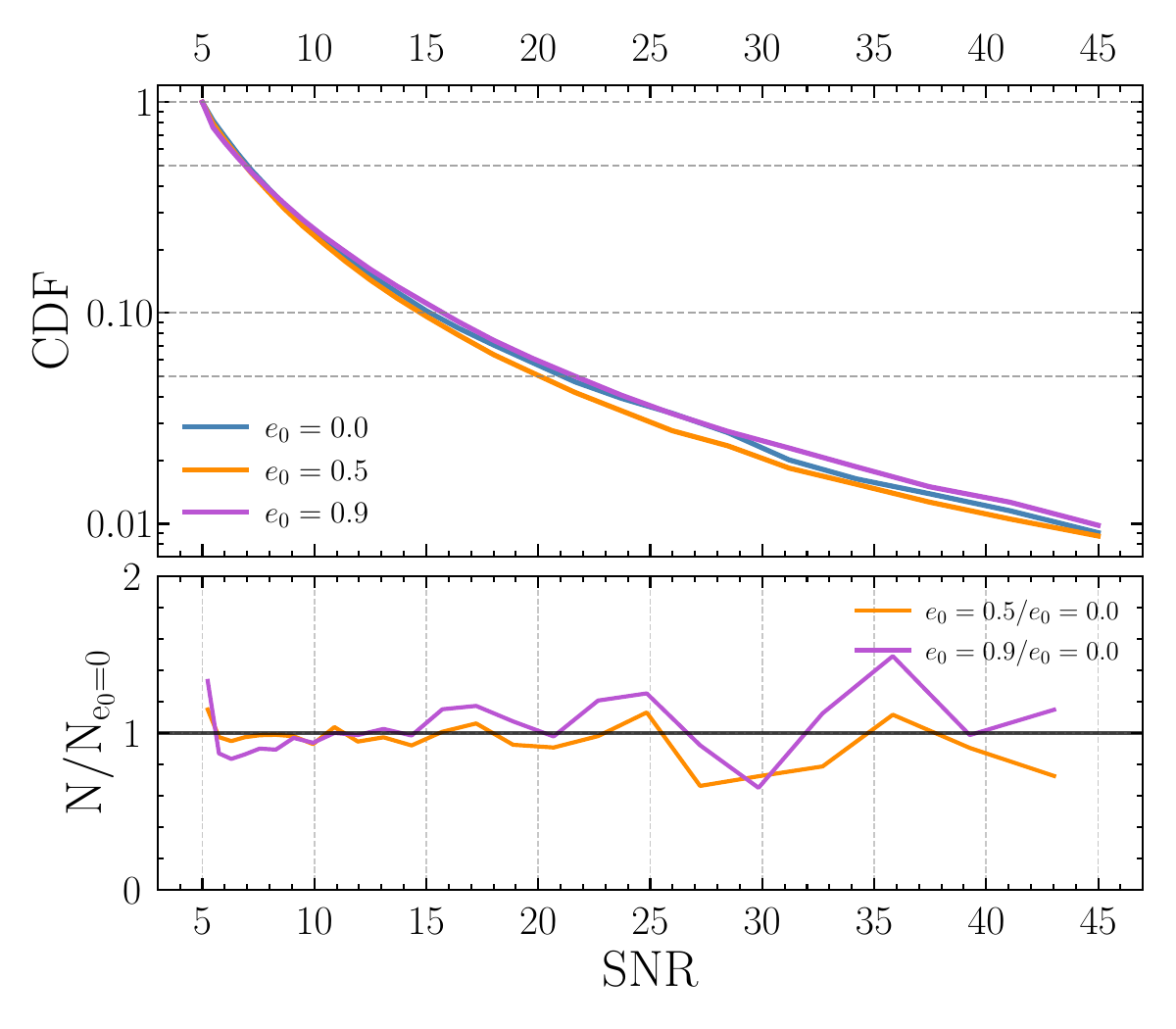}
    \caption{Cumulative distribution function (CDF) of the SNR featured by MBHBs detected by 30-year SKA PTA. Blue, orange, and purple curves represent the CDF of all the models generated with $e_0\,{=}\,0,0.5$ and $0.9$, respectively.  The lower panel presents the number of sources found with a given value of SNR for $e_0\,{=}\,0.5$ and $e_0\,{=}\,0.9$ models (orange and purple line) normalized with the same value found in the $e_0\,{=}\,0$ case.} 
    \label{fig:SNR_distribution}
\end{figure}

\subsection{Number of resolvable sources} \label{sec:N_res}

\begin{figure*}
    \centering
    \includegraphics[width=2\columnwidth]{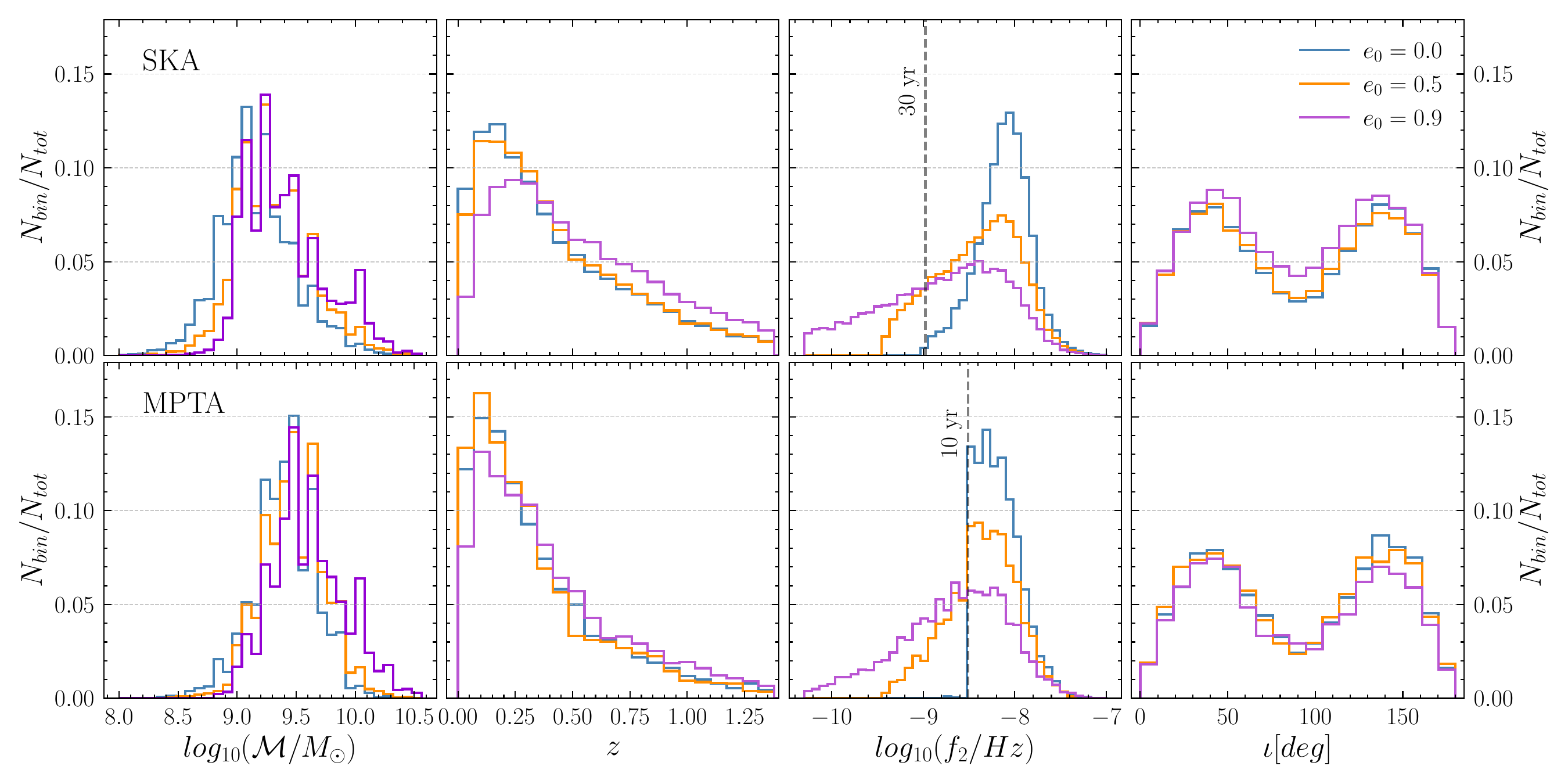}
    \caption{Chirp mass distribution ($\mathcal{M}$, upper left panel),  redshift ($z$, upper right panel), twice the observed Keplerian frequency ($f_2$, lower left panel), and inclination angle ($i$, lower right panel) of the detected MBHBs. Each color represents the distributions when different eccentricity values are used to evolve the MBHB population ($e_0 = 0$, blue, $e_0 = 0.5$, orange and $e_0 = 0.9$ purple). $N_{Bin}$ represents the number of objects in a given bin of the histogram while $N_{tot}$ is the total number of objects analyzed. While the upper panels represent the results for 30-year SKA PTA, the lower ones correspond to 10-year MPTA.}
    \label{fig:Source_Properties}
\end{figure*}

\begin{figure}
    \centering
    \includegraphics[width=1\columnwidth]{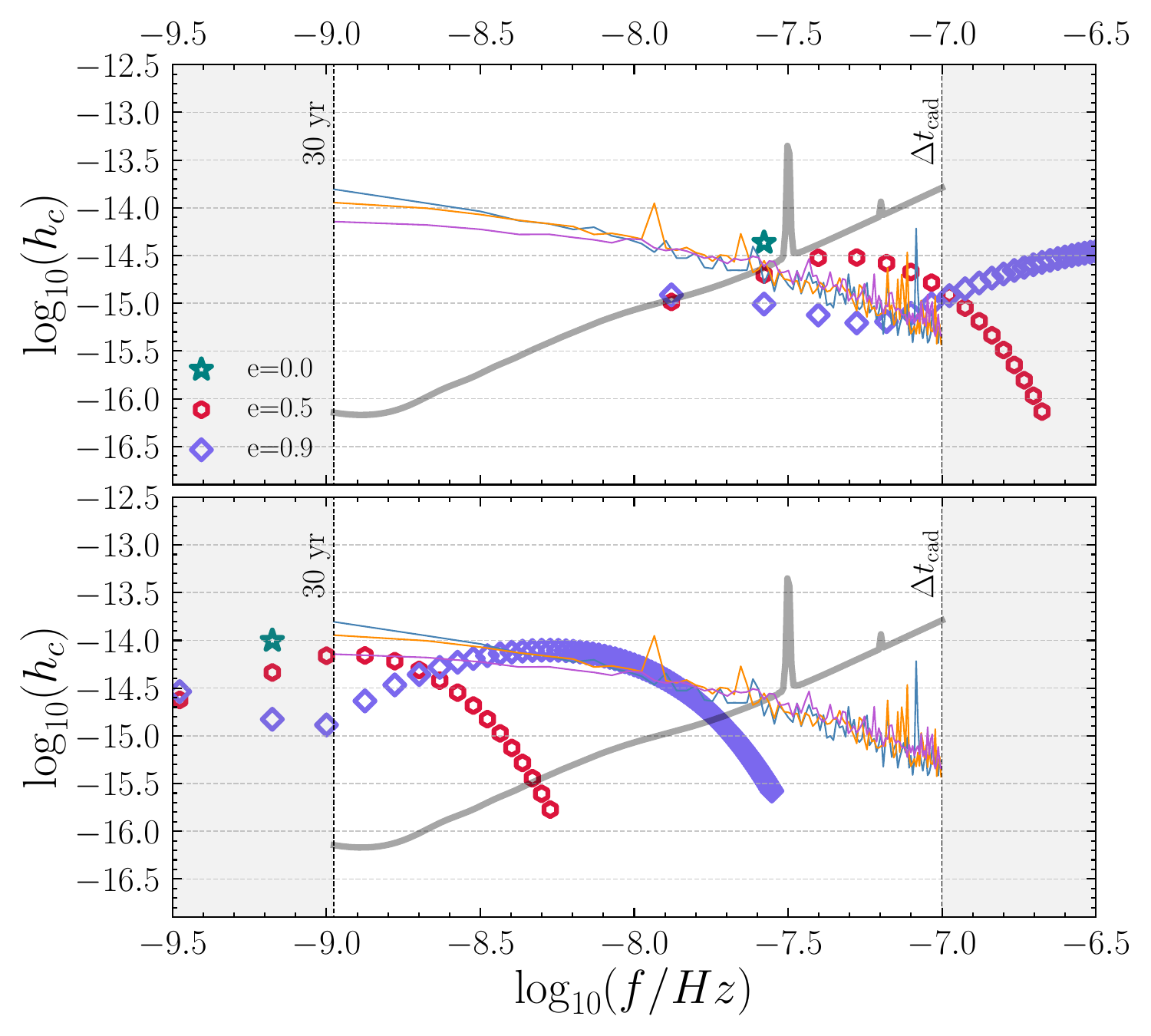}
    \caption{Characteristic strain ($h_c$) as a function of observed frequency ($f$). The upper panel shows a binary  with $\mathcal{M}\,{\sim}\,10^{8.5}\, \rm M_{\odot}$ and the observed Keplerian frequency $f_{k}\,{\sim}\,10^{-8}\, \rm Hz$; while the lower panel shows a binary of  $\mathcal{M}\,{\sim}\,10^{10}\, \rm M_{\odot}$ and $f_{k}\,{\sim}\,10^{-9.5}\, \rm Hz$ outside the SKA PTA sensitivity curve. In each panel,
    blue, orange, and purple lines represent the root mean square of the sGWB generated by all the models with $e_0\,{=}\,0,0.5$ and $0.9$, respectively. The black line corresponds to the sensitivity curve of SKA PTA (white and red noise) 30-year data. The colored dots represent how the signal of an MBHB is distributed across different frequencies when the eccentricity is varied between $0, 0.5$ and $0.9$. }
    \label{fig:Strain_and_Sensitivuty_Curve}
\end{figure}

%The upper panel of Fig.~\ref{fig:Resolvable_Sources_Eccentricity_MeerKAT_SKA} depicts the median number of resolvable sources ($N_{\rm res}$) detected by the SKA and MPTA. To show the impact of red noise Fig.~\ref{fig:Resolvable_Sources_Eccentricity_MeerKAT_SKA} shows the number of resolved MBHB by SKA PTA with (red) and without (orange) considering it. Generally speaking, when red noise is included the number of resolvable sources drops by a fraction  $\sim 30\%$. From now on, if not noticed differently, when we refer to SKA PTA that feature both white and achromatic red noise as described in Section  \ref{sec:Pulsar_Arrays}. The results have been divided according to the eccentricity at which the MBHB population was initialized ($e_0$). This classification allows us to understand the role of the eccentricity of the global MBHB population on the prospects of CGW detection. The median number of resolvable sources for 10-year MPTA is $4$, independently of $e_0$. Conversely, 30-years SKA provides larger $N_{\rm res}$ values ($\,{\sim}\,35$), increasing with eccentricity.\\

The upper panel of Fig.~\ref{fig:Resolvable_Sources_Eccentricity_MeerKAT_SKA} shows the median number of resolvable sources ($N_{\rm res}$) detected by the SKA and MPTA. The results have been divided according to the eccentricity at which the MBHB population was initialized ($e_0$). This classification allows us to understand the role of the eccentricity of the global MBHB population on the prospects of CGW detection. The median number of resolvable sources for 10-year MPTA is $4$, independently of $e_0$. Conversely, 30-years SKA provides larger $N_{\rm res}$ values ($\,{\sim}\,35$), increasing with eccentricity. In particular, the number of detected binaries starts to increase when $e_0\,{>}\,0.2$. This trend can be ascribed to the appearance of resolvable high-eccentric MBHBs with observed Keplerian frequency outside of the PTA frequency range. Given their large eccentricity, these systems can push a large fraction of their GW signal inside the PTA band (more details in the description of Fig.~\ref{fig:Source_Properties} and Fig.~\ref{fig:Strain_and_Sensitivuty_Curve} below). The eccentricity distribution of the detected MBHBs is presented in the lower panel of Fig.~\ref{fig:Resolvable_Sources_Eccentricity_MeerKAT_SKA}. Regardless of the adopted array, the eccentricity distribution of resolved sources peaks at lower values compared to the underlying overall MBHB population. Therefore, the eccentricity of the detected systems is not a good tracer of the eccentricity of the global MBHB population. This is because the more massive binaries, which circularize faster (see Eq.~\ref{eq:eccentricity_Evolution}), are also the more likely to be detected. Compared to MPTA, SKA PTA can generally observe more eccentric MBHBs, which is expected due to its longer timespan. In fact, the SKA PTA sensitivity extends to lower frequencies, where MBHBs had less time to circularize due to GW emission. For completeness, Fig.~\ref{fig:Resolvable_Sources_Eccentricity_MeerKAT_SKA} depicts the number of resolvable sources for SKA PTA when only the pulsar white noise is taken into account. As shown, when the red noise is neglected the number of resolvable sources increases by ${\sim}\,30\%$.\\ 
%since it is sensitive lo
%Resolved binaries in models initialized with $e_0\,{<}\,0.4$ feature $e_{\rm rs}\,{\sim}\,0.2$ with a few systems with higher eccentricity. On the contrary, at $e_0\,{>}\,0.4$ there is a relevant number of resolved MBHBs with $e_{\rm rs}\,{>}\,0.4$ and it coincides with the rise of $N_{\rm res}$. Interestingly, this high-eccentric resolvable population is larger in SKA PTA than in MPTA, explaining in the latter case the no correlation between $N_{\rm res}$ and $e_0$. 

Finally, Fig.~\ref{fig:SNR_distribution} shows the distribution of the SNR  of resolvable sources. For clarity, we only presented the results for the SKA PTA given that MPTA features the same trends (but extended down to $\rm SNR\,{=}\,3$). As we can see, 90\% of the detected systems present $\rm SNR\,{<}\,15$, but there is a large tail towards larger values. Despite being just a few, the remaining 10\% of sources with $\rm SNR\,{>}\,15$ will be optimal targets for multimessenger astronomy, since their sky localization will be small enough to perform electromagnetic follow-ups (see Section~\ref{sec:ParamEstimation} and \citealt{Goldstein2019}). Finally, to compare the SNR distributions for models initialized with different eccentricities we compute the ratio between the number of sources at different SNR bins for $e_0=0.5,0.9$ by the detected population with $e_0=0.0$. As can be seen, no major differences in the SNR distribution are found.

\subsection{Properties of the resolvable sources}

In this section, we study the properties of the resolvable sources and explore possible dependencies with the eccentricity of the underlying MBHB population. For the sake of clarity, the analysis has been done only using three reference eccentricity models: $e_0\,{=}\, 0.0, 0.5$ and $0.9$.\\

The left panels of Fig.~\ref{fig:Source_Properties} present the chirp mass distribution of the MBHB population detected by SKA and MPTA experiments. As shown, both PTAs will detect MBHBs with $\mathcal{M}\,{\sim}\,10^{9.5}\, \rm M_{\odot}$, although MPTA will be biased towards more massive systems given its lower sensitivity. Interestingly, the detection of $\mathcal{M}\,{\lesssim}\,10^{9}\, \rm M_{\odot}$ binaries by SKA is preferred when the underlying MBHB population is initialized with low eccentricities ($e_0\,{<}\,0.5$). This is due to the typical eccentricity and observed Keplerian frequency of $\mathcal{M}\,{<}\,10^9\, \rm M_{\odot}$ systems. These MBHBs are placed at $f_k\,{\sim}\,10^{-8.5}\, \rm Hz$ independently of $e_0$, but their eccentricity raises when $e_0$ increases (e.g. ${\sim}\,0.4$ and ${\sim}\,0.6$ for $e_0\,{=}\,0.5, 0.9$ models, respectively). These relatively high values of $f_k$ and eccentricity cause these systems to emit part of their GW strain at high frequencies where the PTA sensitivity is already degrading. The net effect is the decrease of the source SNR with respect to a non-eccentric case. To illustrate this, the top panel of Fig.~\ref{fig:Strain_and_Sensitivuty_Curve} presents the characteristic GW strain versus observed GW frequency for three binaries with the same mass and Keplerian frequency but different eccentricities. As shown, for circular binaries all the emitted power falls in the frequency region in which the PTA has the best sensitivity. However, for the extreme case of $e\,{>}\,0.5$ most of the power is pushed at $f\,{>}\,3\,{\times}\,10^{-8} \, \rm Hz$ where the PTA is the less sensitive. Consequently, our analysis suggests that the detection of low-mass MBHBs ($\mathcal{M}\,{<}\,10^{9}\, \rm M_{\odot}$) will be hindered in highly eccentric populations.\\

The redshift distribution of the resolvable sources is presented in the middle-left panels of Fig.~\ref{fig:Source_Properties}. The distribution peaks at $z\,{<}\,0.25$, independently of the PTA experiment used and the eccentricity of the underlying MBHB population. The SKA resolved population has a longer tail at high redshifts, due to its better sensitivity.  
Moreover, there is a small trend towards higher redshifts with increasing eccentricity, more prominent in SKA than MPTA.
%Deviation from this trend is seen in highly eccentric cases for SKA where there is a suppression of low-$z$ cases. This is just related to the lack of low mass MBHBs ($\mathcal{M}\,{\lesssim}\,10^{9}\, \rm M_{\odot}$) which in turn are the ones placed at the lowest redshift. 
The frequency distribution of the resolved binaries is shown in the middle-right panels of Fig.~\ref{fig:Source_Properties}. The peak of the distribution seats around ${\sim}10^{-8.5}-10^{-8}\, \rm Hz$, being systematically higher for SKA PTA given its better sensitivity at high frequencies. As anticipated in Section~\ref{sec:N_res}, models with eccentric binaries enable the detection of MBHBs whose $f_2\,{=}\,2f_{k}$ is smaller than the minimum frequency allowed by the PTA observing time. In the extreme case of $e_0\,{=}\,0.9$, up to half of the detected systems display this feature in the MPTA array. An illustrative example of how the strain is distributed among all the harmonics for a source with Keplerian frequency outside the PTA band can be seen in the lower panel of Fig.~\ref{fig:Strain_and_Sensitivuty_Curve}. 
Finally, the inclination distribution shown in the rightmost panels of Fig.~\ref{fig:Source_Properties} is bimodal, preferring face-on/face-off binaries with respect to the observer ($i\,{<}\,50\, \rm deg$ and $i\,{>}\,125\, \rm deg$).  This is simply due to the angular pattern emission of GWs, which are stronger along the binary orbital angular momentum axis. 
%for which the signal is stronger when the binary angular momentum direction is mostly aligned with the line of sight.

\subsection{SMBHB parameter estimation}

\begin{figure*}
    \centering
    \includegraphics[width=2\columnwidth]{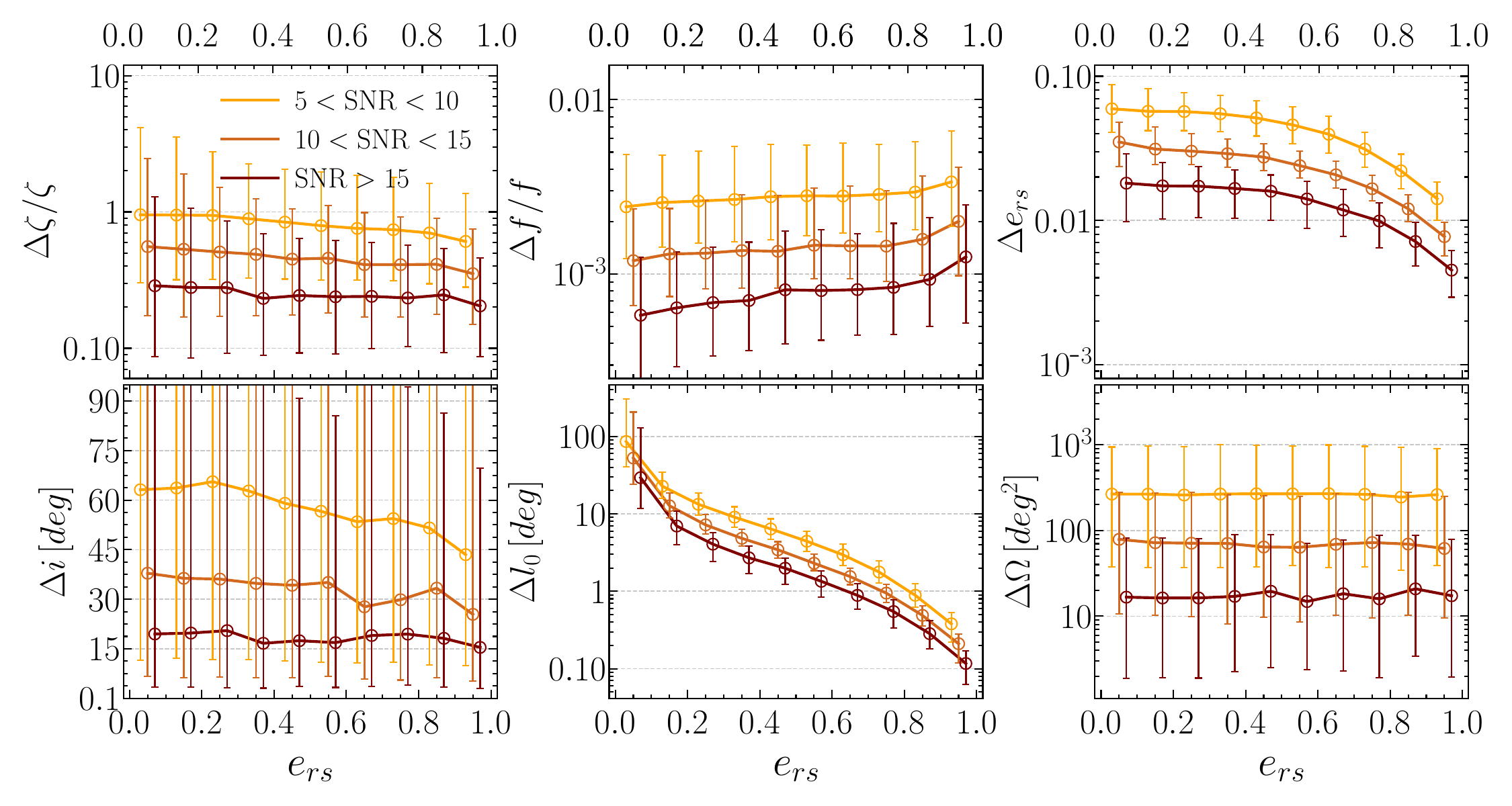}
    \caption{Accuracy in recovering the source parameters as a function of the eccentricity of the detected source, $e_{\rm rs}$: GW amplitude ($\Delta \zeta/\zeta$, top left), Keplerian frequency ($\Delta f/f$, top middle), eccentricity ($\Delta e_{rs}$, top right), inclination angle ($\Delta i$, bottom left), the initial phase of the orbit ($\Delta l_0$, bottom middle) and sky localization ($\Delta \Omega$, bottom right). All the results have been divided into three different SNR bins $\rm 3\,{<}\,SNR\,{<}\,10$ (orange), $\rm 10\,{<}\,SNR\,{<}\,15$ (dark orange) and $\rm SNR\,{>}\,15$ (dark red).}
    \label{fig:Delta_vs_e}
\end{figure*}

Here, we explore the precision to which the CGW source parameters can be determined. To this end, we make use of the procedure presented in Section~\ref{sec:ParamEstimation}. We focus on parameters of astrophysical relevance. Specifically, the GW amplitude, $\zeta$, the observed Keplerian frequency, $f_k$, the orbital eccentricity, $e_{rs}$, the inclination angle, $i$,
%\footnote{Notice that the determination of the inclination angle is relevant for multimessenger studies since, for example, electromagnetic emission in the form of relativistic jets could be aligned with the angular momentum of the binary.},
and the initial orbital phase, $l_0$. The latter parameters might help the identification of distinctive electromagnetic counterparts. In fact, accreting binaries at low inclination angles might appear as Type I AGN, displayng considerable variability in the optical/UV, while relativistic jets could be observable for nearly face-on systems \citep[e.g.][]{2024PhRvD.109j3024F,2024MNRAS.532..506G}. Moreover, precise phase determination of eccentric binaries allows to clearly identify the periastron passage epochs, which can be associated to a dimming in the electromagnetic emission due to temporary mini-disc disruptions caused by the close flyby of the two MBHs \citep{2024arXiv240205175C}. Finally, we combine the two angles defining the source position in the sky to determine the 2D sky localization uncertainty as \citep{SesanaVecchio2010}: 
\begin{equation} \label{eq:Sky_Localization}
    \Delta \Omega \,{=}\,2\pi \sqrt{(\sin \theta \Delta \theta \Delta \phi)^2 - (\sin \theta  \, \sigma_{\theta \phi})^2},%4 \Delta \phi \Delta \theta \cos (\theta)
\end{equation}
where $\sigma_{\theta, \phi}$ is the correlation coefficient between $\theta$ and $\phi$ computed from the Fisher matrix. With this definition the probability of a GW source to be found outside a certain solid angle $\Delta \Omega_0$ is proportional to  $e^{-\Delta \Omega_0/\Delta\Omega}$. Consequently, $\Delta \Omega$ is an important quantity to take into account given that it provides information about the accuracy of pinpointing the GW source in the sky. Moreover, its specific value will shed light on the possibility of carrying out multimessenger studies by placing constraints on the size of the area to scan for electromagnetic follow-ups \citep[see, e.g.,][]{Lops2023,Petrov2024}.\\

Results are presented in Fig.~\ref{fig:Delta_vs_e} for the SKA PTA as a function of the eccentricity of the detected source, $e_{\rm rs}$, to determine its potential impact on the parameter estimation. MPTA parameter estimation features the same trends and is shown in Appendix~\ref{appendix:Recovery_MPTA}.
Since the estimation precision depends on the source SNR, we have performed this exploration at fixed bins of SNR: $\rm 3\,{<}\,SNR\,{<}\,10$, $\rm 10\,{<}\,SNR\,{<}\,15$, and $\rm SNR\,{>}\,15$. For each case, we show the median value on the error of the parameter recovery and the central 68\% of the distribution. The recovery of the GW amplitude displays a small correlation with $e_{rs}$, slightly improving for high eccentric binaries. For instance, the median relative error for systems with $\rm SNR\,{>}\,15$ and $e_{rs}\,{<}\,0.1$ is ${\sim}\, 30\%$ while for high eccentric cases is reduced down to ${\sim} \, 20\%$. As it is the case for all parameters, the GW amplitude recovery precision scales linearly with the inverse of the SNR. Notably, while at $\rm SNR\,{>}\,15$ the source amplitude can be determined with a median relative error of 0.2-0.3, at $\rm SNR\,{<}\,10$ it is poorly constrained. Conversely, the Keplerian frequency is extremely well determined, with a relative error that is always smaller than $1\%$. In this case, the trend with eccentricity is reversed, with the median error increasing with $e_{rs}$. 
%follows a similar trend but is better constrained given that the relative error is already ${\sim}\,1\%$ (${\sim}\,0.1\%$) at $\rm SNR\,{<}\,10$ ($\rm 10\,{<}\,\rm SNR\,{<}\,15$). Concerning the orbital parameters, 
The error associated with the binary eccentricity improves for highly eccentric systems, with values as low as ${\sim}\,1\,{-}\,5\%$ at $e_{rs}\,{>}\,0.6$. The inclination of the MBHB orbit is essentially unconstrained, especially for systems at small SNR and for small eccentricities. 
The initial phase of the orbit displays a clear dependence on $e_{\rm rs}$, being better constrained for large eccentric cases. For instance, at $10\,{<}\,\rm SNR\,{<}\,15$ $\Delta l_0$ associated with  $e_{\rm rs}\,{<}\,0.2$ MBHB is ${\sim}\,10\, \rm deg$ while it drops down to ${\sim}\,1\, \rm deg$ for $e_{\rm rs}\,{>}\,0.6$. This is not surprising since for an eccentric orbit GW emission is strongly localized close to the pericenter, allowing a precise measurement of the orbital phase of the system.

%the phasing of the observed signal strongly depends on $l_0$. Conversely, $l_0$ is not even defined for a circular orbit.\\
%The direction of the pericenter displays a dependence on $e_{\rm rs}$, being better constrained for large eccentric cases. For instance, at $10\,{<}\,\rm SNR\,{<}\,15$ $\Delta \gamma$ associated with  $e_{\rm rs}\,{<}\,0.2$ MBHB is ${\sim}\,100\, \rm deg$ (i.e. essentially undetermined), while it drops down to ${\sim}\,20\, \rm deg$ for $e_{\rm rs}\,{>}\,0.6$. This is not surprising since the direction to the periastron is a well defined parameter only for an eccentric binary, and it has a stronger influence on the waveform shape for higher eccentricities.

%becomes better for an eccentric orbit GW emission is strongly localized close to the pericenter, and therefore the phasing of the observed signal strongly depends on $l_0$. Conversely, $l_0$ is not even defined for a circular orbit. %\as{Sorry I got confused on this parameter. I thought $l_0$ was the direction of the periastron and not the initial phase of the orbit. What's written here makes sense for the direction of the periastron since it is better defined and measurable for high eccentricity, but not really for the initial phase. I would expect the initial phase of the orbit to be quite measurable for all eccentricities. Can you please check the estimate of both $l_0$ and $\gamma$?}\\

Finally, the lower right panel of Fig.~\ref{fig:Delta_vs_e}  presents the sky-localization. Interestingly, it does not show any dependence with $e_{\rm rs}$ but, as expected, it strongly improves with SNR, since the parameter has a theoretical scaling with $\rm SNR^{-2}$. Binaries detected at $\rm 5 \, {<} \, SNR \,{<} \, 10$ have a median sky-localization of $\Delta \Omega \,{\sim}\,200\, \rm deg^2$, making multimessenger follow-ups extremely challenging. On the other hand, systems with $\rm SNR\,{>}\,15$ feature median $\Delta \Omega\,{\sim}\,20\, \rm deg^2$. Note that the 68\% confidence region extends down to ${\approx}\, 2 \, \rm deg^2$. Since SKA can resolve 30-40 binaries and about 10\% of them will have SNR$>15$ (see Fig.~\ref{fig:SNR_distribution}), we can therefore expect at least one CGW with source localization at the $\sim $deg$^2$ level, which would be a perfect target for electromagnetic follow-ups. 

\begin{figure}
    \centering
    \includegraphics[width=1\columnwidth]{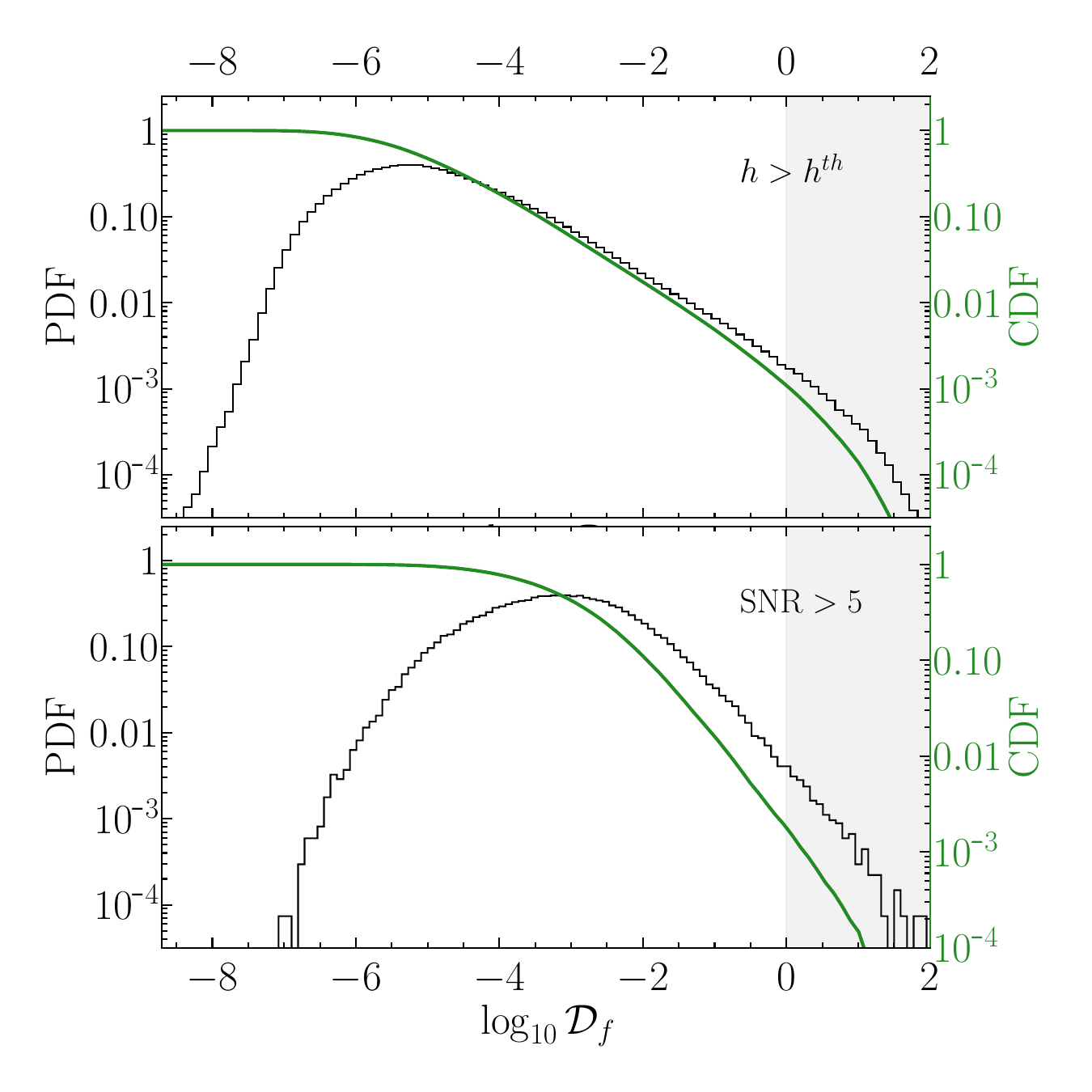}
    \caption{Probability distribution function (PDF) of $\mathcal{D}_f$ for binaries featuring $h\,{>}\,h^{\rm th}$ (top panel) and binaries detected by SKA PTA ($\rm SNR\,{>}\,5$, bottom panel). In both panels, green lines correspond to the cumulative distribution function (CDF).}
    \label{fig:Ev_f}
\end{figure}

\section{Caveats} \label{sec:Caveats}
In this section, we discuss the main caveats and assumptions related to the methodology. 
%used to detect continuous gravitational wave sources from the nHz sGWB. 

\subsection{Time evolving binaries}

We have assumed that the MBHB orbital frequency does not evolve during the PTA observation time. However, this simplification may not hold, especially for massive and high-frequency binaries given their shorter GW timescales (see Eq.~\ref{eq:frequency_Evolution}). To explore the fraction of MBHBs in our catalogues in which the non-evolving assumption is not fulfilled, we have computed the following quantity:
\begin{equation}
    \mathcal{D}_f \,{=}\, \frac{\left[\frac{df_k}{dt} T_{obs}\right]}{\Delta f}.
\end{equation}
Here $df_k/dt$ is determined according to Eq.~\eqref{eq:frequency_Evolution} but for simplicity accounting only the GW term, while the factor $(df_k/dt)\,{\times}\,T_{obs}$ corresponds to the variation of the observed Keplerian frequency over the PTA observational time. The division by $\Delta f$ accounts for the total variation of the MBHB frequency within the frequency bin width given by the PTA observation span. The upper panel of Fig.~\ref{fig:Ev_f} presents the distribution of $ \mathcal{D}_f $ for all the binaries in our catalogue whose $h\,{>}\,h^{\rm th}$ (see Section~\ref{sec:ResolvingBinaries}). As shown, the distribution peaks at low values of $ \mathcal{D}_f $ (${\sim}\,10^{-6}$), implying an almost null evolution of the binary frequency. Nevertheless, some cases show $ \mathcal{D}_f\,{>}\,1$, but they correspond to less than 0.1\% of the MBHB population. The lower panel of Fig.~\ref{fig:Ev_f} presents the $\mathcal{D}_f$ distribution only for the subset of sources that are resolvable by SKA PTA. Not surprisingly, the distribution for this sub-sample of binaries peaks at larger values. This shift is caused by the fact that individually resolvable MBHBs are intrinsically systems with large masses and high frequencies (see Fig.~\ref{fig:Source_Properties}). Despite this, the bulk of the system feature $ \mathcal{D}_f \,{\sim}\,10^{-3}$, consistent again with non-evolving binaries. Also for this sub-sample, binaries with $ \mathcal{D}_f\,{>}\,1$ only account for 0.1\% of the resolvable sources. In light of these results, we can conclude that our assumption of non-evolving binaries can be safely adopted. \\

\subsection{Pericenter precession}

\begin{figure}
    \centering
    \includegraphics[width=1\columnwidth]{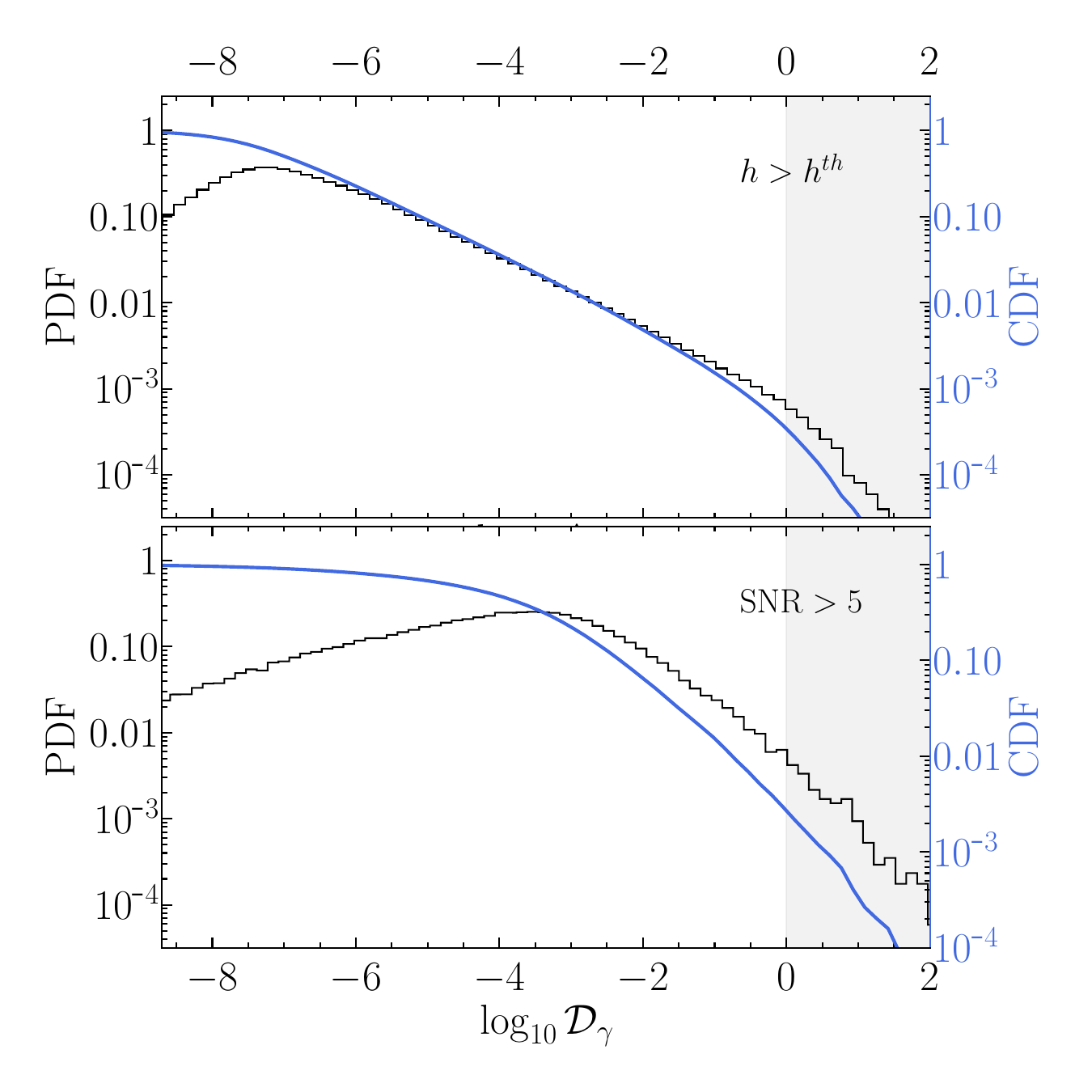}
    \caption{Probability distribution function (PDF) of $\mathcal{D}_{\gamma}$ for binaries featuring $h\,{>}\,h^{\rm th}$ (top panel) and binaries detected by SKA PTA ($\rm SNR\,{>}\,5$, bottom panel). In both panels, blue lines correspond to the cumulative distribution function (CDF).}
    \label{fig:Ev_Pericenter}
\end{figure}

Another assumption that is relevant to discuss concerns the inclusion of the pericenter precession. To quantify its impact on the SNR recovery, we adopted the same criteria presented in \cite{SesanaVecchio2010}. The precession of the pericenter induces an additional shift in the observed Keplerian frequency given by $f_k + \dot{\gamma}/\pi$ with
\begin{equation} \label{Eq::pericenter preception}
\dot{\gamma}\,{=}\, \dfrac{d\gamma}{dt} = 6 \pi f_k \dfrac{ (2\pi f_k (1+z) M G)^{2/3} }{(1-e^2)c^2}.
\end{equation}
This causes a bias in the recovery of the orbital frequency. However, this effect can be neglected as long as the shift caused by pericenter precession over the observed time is small compared to the frequency resolution of the detector, $\Delta f$. This is equivalent to enforce the condition $\mathcal{D}_\gamma\,{<<}\,1$, where:
\begin{equation}
    \mathcal{D}_\gamma \,{=}\, \frac{\left[\frac{d^2\gamma}{dt^2} T_{obs}\right]}{\Delta f},
\end{equation}
with:
\begin{equation}
   \dfrac{d^2\gamma}{dt^2} \,{=}\, \dfrac{96 (2\pi)^{13/3}}{(1-e^2)c^7} [(1+z)M]^{2/3}\mathcal{M}_z^{5/3}G^{7/3}f_k^{13/3}.
\end{equation}
Fig.~\ref{fig:Ev_Pericenter} shows the distribution of $\mathcal{D}_\gamma$ for all MBHBs with $h\,{>}\,h^{th}$ (top panel) and for those detected by the SKA PTA  (lower panel). Similar to the $\mathcal{D}_f$ result, the systems with $\mathcal{D}_\gamma\,{>}\,1$ represent only 0.1\% of the resolved MBHBs. As a consequence, the effect of the pericenter precession can be ignored for our astrophysical-motivated populations of MBHBs.

\subsection{Pulsar term}
Similarly to other works, we did not account for the pulsar term since its inclusion in the matched filtering methodology requires a precise estimate of the distance between the pulsar and the Earth. Up to date, only a very reduced sample of pulsars has such an accurate measurement of this quantity. %given the fact that even small variations in the pulsar distance led to inefficient parameter estimation recovery (\citealp{2016ApJ...817...70T}, \citealp{Ellis2014}). 
Despite this, ongoing efforts are being made to calculate the pulsar distance via the measurement of pulsar spin down and annual parallax motion \citep[see e.g][]{Reardon2016}. 

When it comes to 2D sky localization, including the Pulsar term in the analysis does not appear to make a significant difference in the size of the localization area, at least in the case of circular, GW-driven binaries. It can, however, cause a small bias compared to the true sky location (Ferranti et al. in prep.). Therefore,  while Earth term-only estimates are robust in terms of localization precision, including the pulsar term in the analysis might be required to pinpoint the correct direction in the sky. 

Including the pulsar term in the CGW searches could also provide key information on how the MBHB evolves during the times needed for the pulse to cover the Earth-Pulsar distance. Under the assumption of GW-driven binaries, identification of the pulsar term allows to effectively separate the system chirp mass from the distance in the signal amplitude parameter $\zeta$ (Ferranti et al. in prep for examples involving circular binaries), greatly improving 3D localization of the source in the sky. This assumption, however, is not necessarily fulfilled. At the low frequencies probed by PTAs, MBHBs can be still coupled to their environment, especially at the time of Pulsar-term production. In fact, since the typical Earth-Pulsar separation can range up to thousands of light years, the Pulsar and Earth terms of the signal could inform us about the binary properties at two different evolutionary stages. In this case, the change of parameters such as the orbital frequency and the eccentricity could help to better understand the environment in which the MBHB resided.  % since the main process that rules the MBHB dynamic led to different evolutionary paths. 
As shown in Eq.~\eqref{eq:frequency_Evolution}, the Keplerian frequency of a binary evolving only due to GW emission varies as ${\propto}\,f^{11/3}$, while if its dynamics is ruled by stellar scattering events, it changes as ${\propto}\, f^{1/3}$. Including environmental coupling, however, increases the number of parameters in the model, and whether GW and environmental effects can be efficiently separated in a real analysis has still to be investigated. 
%For these reasons, in future work, we aim to include the pulsar term in the computation of the SNR.

\subsection{Further complications in source detectability}
Throughout this work, we used a simple SNR criterion to define source detectability, regardless of the nature of the GW signal. However, the shape of the waveform can be significantly different for circular and highly eccentric binaries and while the detectability of the former has been extensively demonstrated in the literature \citep{2012PhRvD..85d4034B,Ellis2012,Ellis2014}, much less has been done on the eccentric binary front \citep{2016ApJ...817...70T}. This is especially true for sources with $f_2<1/T$, which can constitute up to 50\% of the resolvable CGWs in the limit of high eccentricities for the MPTA (see Fig.~\ref{fig:Source_Properties}). For these systems, the waveform consists of a single burst-like spike coincident with the binary periastron passage \citep[e.g.][]{2010MNRAS.402.2308A}, which is very different from a repeated sinusoidal pattern. Although analytical templates can certainly be constructed for such signals, the effectiveness of match filtering in extracting them from real data has still to be investigated. 

\section{Conclusions} \label{sec:Conclusions}
In this work, we studied the capability of future PTA experiments of detecting single MBHBs under the natural assumption that the sGWB is produced by an eccentric MBHB population. To this end, we have generalized the standard approach used in PTA to assess the observability of circular MBHBs, by computing the SNR and Fisher Information Matrix for eccentric systems. We have adopted a 10-year MPTA and 30-year SKA PTAs and applied our analysis to a wide number of simulated eccentric MBHB populations, compatible with the latest measured amplitude of the sGWB. The main results can be summarized as follows:

 \begin{itemize}
    \item The expected number of resolvable sources detected by a 10-year MPTA ($\rm SNR\,{>}\,3$) is $4^{+3}_{-2}$ (68\% credible interval) %\as{Check that the number is correct} 
    with no dependence on the eccentricity of the underlying MBHB population.\\ 
    
    \item The extraordinary sensitivity of a 30-year SKA PTA will enable the detection of $30^{+11}_{-10}$ (68\% credible interval) sources with $\rm{SNR}>5$ for initially circular MBHB population. This number grows to $40^{+15}_{-15}$ in the case of very high MBHB initial eccentricity ($e_0\,{=}\,0.9$). This is mostly caused by highly eccentric binaries with Keplerian frequency ${\lesssim}\,10^{9}\, \rm Hz$ pushing part of their power into the SKA sensitivity band.\\

    \item The resolved MBHBs do not follow the eccentricity distribution of the underlying MBHB population. Instead, they tend to favor lower eccentricities. This is caused by the fact that the bulk of the detected MBHB population is placed in the frequency range $10^{-8.5}\,{-}\,10^{-8}\, \rm Hz$. At those frequencies, GW emission is expected to dominate and, as a consequence, partial circularization of the binary orbit has already taken place. Practically, this means that massive and high-frequency systems, most likely to be detected, should display low eccentricities with respect to the bulk of the population.\\

    \item The chirp mass ($\mathcal{M}$) of the resolvable sources is ${\gtrsim}\,10^{8.5}\,\rm M_{\odot}$, but it depends on the specific PTA experiment. While the median value for MPTA is ${\sim}\,10^{9.5}\,\rm M_{\odot}$, that for SKA shifts down to ${\sim}\,10^{9}\,\rm M_{\odot}$. The results also show that the detection of binaries with $\mathcal{M}\,{\lesssim}\,10^{9}\, \rm M_{\odot}$ is strongly disfavored, especially when the eccentricity of the underlying MBHB population is large.\\ 

    \item The distribution of resolvable sources peaks at $z\,{<}\,0.25$, regardless of the PTA used but, unsurprisingly, it is more skewed towards low-$z$ for MPTA. Their typical frequency at the second harmonic ($f_2$) sits at ${\sim}\,10^{-8.5}\, \rm Hz$ for SKA PTA and increases to ${\sim}\,10^{-8}\, \rm Hz$ for MPTA. The eccentricity of the MBHB population shifts the $f_2$ median value towards low frequencies. %\as{Is this true? By looking at fig 6 I would say that it also shifts the median at lower values}. 
    This is caused by the fact that highly eccentric populations have a significant number of resolvable sources with $f_2 \,{<}\, 1/T_{\rm obs}$. Finally, the inclination between the MBHB and the observer shows a bi-modal distribution with maximum probability for face-one configurations. No correlation with the eccentricity of the MBHB population is seen. \\

    \item The accuracy of recovering the source properties shows a mild dependence on the eccentricity of the system. Whereas the frequency, amplitude, and orbital inclination are almost independent of it, the eccentricity and initial orbital phase of the MBHB orbit show a clear trend. Specifically (and unsurprisingly), these parameters are better constrained for sources with large eccentricities.  \\

    \item The sky-localization does not show any dependence on the MBHB eccentricity. However, it roughly follows the expected $\rm SNR^{-2}$ trend. In particular, binaries detected with $\rm 5 \, {<} \, SNR \,{<} \, 10$ feature a median $\Delta \Omega \,{\sim}\,200\, \rm deg^2$, hindering any possible multimessenger follow-up. MBHBs with $\rm SNR\,{>}\,15$ display a median $\Delta \Omega\,{\sim}\,20\, \rm deg^2$. We note that the scatter around these median values is up to 1dex (68\% confidence), due to the anisotropy of the pulsar distribution and intrinsic properties of the MBHB population. In the most optimistic case, we can expect 30-yr SKA to localize a particularly loud MBHB at a $\sim$deg$^2$ accuracy.  
    
\end{itemize}

In this work, we developed a theoretical framework to assess the detectability and parameter extraction of eccentric MBHB from realistic populations. This allowed us to investigate the performance of future radio facilities such as MPTA and SKA. Being able to fully detect the presence of single MBHB sources at nHz frequencies will be fundamental in determining the astrophysical or cosmological nature of the signal recently reported by worldwide PTAs and will open the era of multimessenger astronomy with MBHBs. In a future work, we plan to implement the procedure presented here in the populations of galaxies, MBHs and MBHBs generated by galaxy formation models to explore the capabilities of associating CGW sources detected by PTAs with galaxies and AGNs.

%-------------------------------------------------------------------------------------------------------------------------------------------
%-------------------------------------------------------------------------------------------------------------------------------------------

\begin{acknowledgements}
       We thank the B-Massive group at Milano-Bicocca University for useful discussions and comments. R.T., D.I.V., A.S. and G.M.S acknowledge the financial support provided under the European Union’s H2020 ERC Consolidator Grant ``Binary Massive Black Hole Astrophysics'' (B Massive, Grant Agreement: 818691). M.B. acknowledges support provided by MUR under grant ``PNRR - Missione 4 Istruzione e Ricerca - Componente 2 Dalla Ricerca all'Impresa - Investimento 1.2 Finanziamento di progetti presentati da giovani ricercatori ID:SOE\_0163'' and by University of Milano-Bicocca under grant ``2022-NAZ-0482/B''.
\end{acknowledgements}


\bibliographystyle{aa} 
\bibliography{references}

\appendix

\section{Accuracy in the recovery of SMBHB parameters for the MPTA case} \label{appendix:Recovery_MPTA}

In this section, we investigate the accuracy of recovering the binary parameters when using the 10-year MPTA. We point out that at very high eccentricities the results are noisy given that in MPTA these types of binaries are rarer than in SKA PTA (see lower panel of Fig~\ref{fig:Resolvable_Sources_Eccentricity_MeerKAT_SKA}). 
For these reasons, the results above $e_{\rm rs}\,{>}\,0.8$ are affected by low statistics and should be taken with caution.\\

\begin{figure*} 
    \centering
    \includegraphics[width=2\columnwidth]{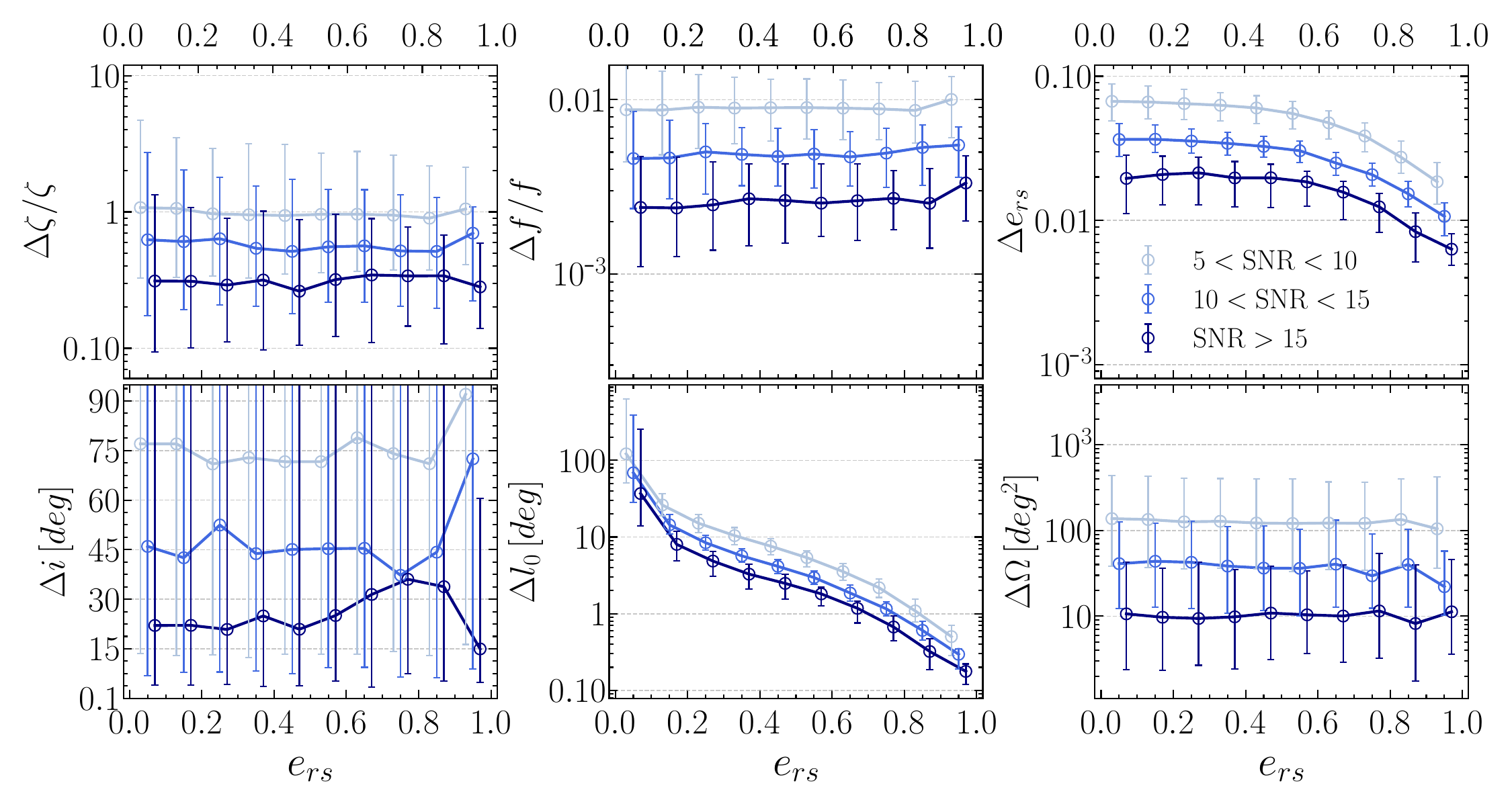}
    \caption{Accuracy for 10-yr MPTA in recovering the source parameters as a function of the eccentricity of the detected source, $e_{\rm rs}$: GW amplitude ($\Delta \zeta/\zeta$, top left), Keplerian frequency ($\Delta f/f$, top middle), eccentricity ($\Delta e_{rs}$, top right), inclination angle ($\Delta i$, bottom left), the initial phase of the orbit ($\Delta l_0$, bottom middle) and sky localization ($\Delta \Omega$, bottom right). All the results have been divided into three different SNR bins $\rm 3\,{<}\,SNR\,{<}\,10$ (pale blue), $\rm 10\,{<}\,SNR\,{<}\,15$ (blue) and $\rm SNR\,{>}\,15$ (dark blue).}
    \label{fig:Delta_vs_e_MPTA}
\end{figure*}

Fig.~\ref{fig:Delta_vs_e_MPTA} presents the results. As shown, the errors in the parameter estimation for the 10-year MPTA generally follow the same trend as the ones for SKA. This confirms that the main driving factor in parameter estimation is the signal-to-noise ratio. Notably, the error in the frequency is worse in MPTA than in SKA at a fixed SNR. This is because the frequency resolution of the PTA is set by the observation time $T_{obs}$. The second interesting result regards the better estimation of the source sky location for MPTA with respect to SKA. This is because the sky position of the MPTA pulsars follows a more isotropic distribution.  Hence, is able to better triangulate the GW source sky position. On the contrary for SKA PTA we select an ultra-realistic pulsar sky distribution, and hence most of the pulsars are located inside the Galatic plane. This highlights the need to choose a distribution of pulsars in the sky as isotropic as possible.
%In future work, we aim to further investigate how the observing and the pulsar sky localization affect the parameter estimation. 

\end{document}